\documentclass[12pt]{article}
\pdfoutput=1
\usepackage{amsfonts,hyperref}
\usepackage{color}
\usepackage{appendix}
\usepackage{amsmath}
\usepackage{graphicx}
\usepackage[latin1]{inputenc}
\usepackage[french,english]{babel}
\usepackage{geometry}
\usepackage{etoolbox}
\usepackage{fancyhdr}
\usepackage{url}
\usepackage{dsfont}
\usepackage{stmaryrd}
\patchcmd{\thebibliography}{\section*}{\section}{}{}

\addto\captionsfrench{}
\addto\captionsenglish{}

\newcommand\encadremath[1]{\vbox{\hrule\hbox{\vrule\kern8pt
\vbox{\kern8pt \hbox{$\displaystyle #1$}\kern8pt}
\kern8pt\vrule}\hrule}}
\def\enca#1{\vbox{\hrule\hbox{
\vrule\kern8pt\vbox{\kern8pt \hbox{$\displaystyle #1$}
\kern8pt} \kern8pt\vrule}\hrule}}

\newcommand\figureframex[3]{
\begin{figure}[bth]
\hrule\hbox{\vrule\kern8pt
\vbox{\kern8pt \vbox{
\begin{center}
{\mbox{\epsfxsize=#1.truecm\epsfbox{#2}}}
\end{center}
\caption{#3}
}\kern8pt}
\kern8pt\vrule}\hrule
\end{figure}
}

\makeatletter
\@addtoreset{equation}{section}
\makeatother
\newtheorem{theorem}{Theorem}[section]
\newtheorem{conjecture}{Conjecture}[section]
\newtheorem{remark}{Remark}[section]
\newtheorem{proposition}{Proposition}[section]
\newtheorem{lemma}{Lemma}[section]
\newtheorem{corollary}{Corollary}[section]
\newtheorem{definition}{Definition}[section]

\def\br{\begin{remark}\rm\small}
\def\er{\end{remark}}
\def\bt{\begin{theorem}}
\def\et{\end{theorem}}
\def\bd{\begin{definition}}
\def\ed{\end{definition}}
\def\bp{\begin{proposition}}
\def\ep{\end{proposition}}
\def\bl{\begin{lemma}}
\def\el{\end{lemma}}
\def\bc{\begin{corollary}}
\def\ec{\end{corollary}}
\def\beaq{\begin{eqnarray}}
\def\eeaq{\end{eqnarray}}
\newcommand{\proof}[1]{{\noindent \bf proof:}\par
{#1} $\square$}

\newcommand{\beq}{\begin{equation}}
\newcommand{\eeq}{\end{equation}}
\newcommand{\bea}{\begin{eqnarray}}
\newcommand{\eea}{\end{eqnarray}}
\newcommand{\beqq}{\begin{equation*}}
\newcommand{\eeqq}{\end{equation*}}
\newcommand{\beaa}{\begin{eqnarray*}}
\newcommand{\eeaa}{\end{eqnarray*}}

\newcommand{\td}[1]{{\tilde{#1}}}

\newcommand{\Pint}{{\int\kern -1.em -\kern-.25em}}

\newcommand{\genus}{{\mathfrak{g}}}

\newcommand\Res{\mathop{{\rm Res}}}

\begin{document}

\sloppy

\pagestyle{empty}
\vspace{10pt}
\begin{center}
{\large \bf {Asymptotic expansions of some Toeplitz determinants via the topological recursion}}
\end{center}
\vspace{3pt}
\begin{center}
\textbf{O. Marchal$^\dagger$}
\end{center}
\vspace{15pt}

$^\dagger$ \textit{Universit\'{e} de Lyon, CNRS UMR 5208, Universit\'{e} Jean Monnet, Institut Camille Jordan, France}
\footnote{olivier.marchal@univ-st-etienne.fr}

\vspace{30pt}

{\bf Abstract}: In this article, we study the large $n$ asymptotic expansions of $n\times n$ Toeplitz determinants whose symbols are indicator functions of unions of arc-intervals of the unit circle. In particular, we use a Hermitian matrix model reformulation of the problem to provide a rigorous derivation of the general form of the large $n$ expansion when the symbol is an indicator function of either a single arc-interval or several arc-intervals with a discrete rotational symmetry. Moreover, we prove that the coefficients in the expansions can be reconstructed, up to some constants, from the Eynard-Orantin topological recursion applied to some explicit spectral curves. In addition, when the symbol is an indicator function of a single arc-interval, we provide the corresponding normalizing constants using a Selberg integral and thus we obtain a generalization of Widom's result.

\vspace{15pt}
\pagestyle{plain}
\setcounter{page}{1}


\section{Introduction: General setting and several reformulations of the problem}
\subsection{General setting}
In this article we are interested in the computation of Toeplitz integrals of the form:
\beq \label{Zn} Z_n(\mathcal{I})=\frac{1}{(2\pi)^nn!}\int_{\mathcal{I}^n} d\theta_1\dots d\theta_n  \prod_{1\leq p<q\leq n}\left|e^{i \theta_p}-e^{i\theta_q}\right|^2\eeq
where $\mathcal{I}$ is a union of $d$ ($d\geq 1$) intervals in $[-\pi,\pi]$:
\beq \mathcal{I}=\bigcup_{j=1}^d \left[\alpha_j,\beta_j\right] \text{  with  } -\pi\leq \alpha_j\leq \beta_j\leq \pi\eeq
Note that in the case of a full support $\mathcal{I}=[-\pi,\pi]$ it is well known that (See \cite{MehtaBook}):
\beq \label{Normalization} Z_n([-\pi,\pi])=1\eeq
Thus, in the rest of the article, \textbf{we assume that} $\boldsymbol{\mathcal{I}}\boldsymbol{\neq} \boldsymbol{[}\boldsymbol{-}\boldsymbol{\pi}\boldsymbol{,}\boldsymbol{\pi}\boldsymbol{]}$. Since the integral is obviously invariant under a global angular translation ($\boldsymbol{\theta}\mapsto \boldsymbol{\theta}-\text{Cste }\mathbf{1}$), we may assume that \textbf{no interval} $\left(\boldsymbol{[}\boldsymbol{\alpha}_\mathbf{j},\boldsymbol{\beta}_\mathbf{j}\boldsymbol{]}\right)_{\mathbf{1}\boldsymbol{\leq}\mathbf{ j}\boldsymbol{\leq}\mathbf{ d}}$ \textbf{contains} $\boldsymbol{\pm}\boldsymbol{\pi}$. Integrals of type \eqref{Zn} can also be understood as the partition functions of a gas of particles restricted to the set $\mathcal{T}=\{e^{it}, t\in \mathcal{I}\}$ with interactions given by $\underset{1\leq p<q\leq n}{\prod}\left|e^{i \theta_p}-e^{i\theta_q}\right|^2$. Moreover, it is also well known that integrals of type \eqref{Zn} are Toeplitz integrals that can be reformulated as the determinant of a $n\times n$ Toeplitz matrix with a non-vanishing symbol (defined below) on some arc-intervals of the unit circle. Asymptotic expansions of Toeplitz determinants and integrals of the form \eqref{Zn} have been studied for a long time and many results already exist in the literature using different strategies. For example, studies using orthogonal polynomials on the unit circle, properties of powers of random unitary matrices, Fredholm determinants, Riemann-Hilbert problems, etc. have been used to tackle the problem. A non-exhaustive list of articles on the subject is \cite{Szego,Widom,Widom,BS,BW,FishHart,BM,BO,DJ,DIZ,Krasovsky,DIK,DIK2,DIK3,RelativeToeplitz}. 

The purpose of this article is to provide a rigorous large $n$ expansion of Toeplitz integrals of the form \eqref{Zn} \textbf{to all order in }$\frac{\mathbf{1}}{\mathbf{n}}$ using the Eynard-Orantin topological recursion defined in \cite{EO}. In particular we shall prove the following results:
\begin{itemize}\item A complete large $n$ expansion when the support is restricted to only one interval (i.e. $d=1$) in Section \ref{Section1Int}.
\item Results up to $O(1)$ when the support is composed of $d=2r+1\geq 3$ intervals of the form $[\alpha_j,\beta_j]=\left[\frac{2\pi j}{2r+1}-\frac{\pi\epsilon}{2r+1},\frac{2\pi j}{2r+1}+\frac{\pi\epsilon}{2r+1}\right]$ with $-r\leq j\leq r$ and $0<\epsilon<1$ in Section \ref{SectionOddd}.
\item Results up to $O(1)$ when the support is composed of $d=2s\geq 2$ intervals of the form $[\alpha_j,\beta_j]=\left[\frac{\pi\left(j-\frac{1}{2}\right)}{s}-\frac{\pi\epsilon}{2s},\frac{\pi\left(j-\frac{1}{2}\right)}{s}+\frac{\pi\epsilon}{2s}\right]$ with $-(s-1)\leq j\leq s$ and $0<\epsilon<1$ in Section \ref{SectionEvenCase}.
\end{itemize}

The strategy used in this article is to reformulate the Toeplitz integrals in terms of some Hermitian matrix integrals (with some restrictions on the eigenvalues support). Then we compute the associated spectral curve and the corresponding limiting eigenvalues density. Using the theory developed in \cite{BG,BG2,BorotGuionnetKoz} we are able to rigorously prove the general form of the large $n$ expansions of the correlators and of the partition function, as well as relate them with quantities computed from the topological recursion. In the case of a single interval, we finally use a Selberg integral to fix the normalization issues of the partition function and thus provide the complete large $n$ expansion of the Toeplitz integrals. We eventually compare our theoretical predictions with numeric simulations performed on the Toeplitz determinant reformulation (which is very convenient for numeric computations) up to $o\left(\frac{1}{n^4}\right)$. 

\subsection{Various reformulations of the problem}

There are several useful rewritings of the integral \eqref{Zn}. We list them in the following theorem:

\begin{lemma}[Various reformulations of the problem]\label{TheoReform} Defining $\mathcal{I}=\underset{j=1}{\overset{d}{\bigcup}} [\alpha_j,\beta_j] \subset(-\pi,\pi)$,  $\mathcal{T}=\underset{j=1}{\overset{d}{\bigcup}} \left\{e^{it}\,,\,t\in [\alpha_j,\beta_j]\right\}$ and $\mathcal{J}=\underset{j=1}{\overset{d}{\bigcup}} [\tan \frac{\alpha_j}{2},\tan\frac{\beta_j}{2}]$, the following quantities are equal to each other:
\begin{enumerate}\item A Toeplitz integral with symbol $f=\mathds{1}_{\mathcal{T}}$:
\bea \label{IntExp} Z_n(\mathcal{I})&=&\frac{1}{(2\pi)^nn!}\int_{[-\pi,\pi]^n} d\theta_1\dots d\theta_n \left(\prod_{k=1}^n f(e^{i\theta_k})\right)  \prod_{1\leq p<q\leq n}\left|e^{i \theta_p}-e^{i\theta_q}\right|^2\cr
&=&\frac{1}{(2\pi)^nn!}\int_{\mathcal{I}^n} d\theta_1\dots d\theta_n  \prod_{1\leq p<q\leq n}\left|e^{i \theta_p}-e^{i\theta_q}\right|^2
\eea
\item The determinant of a $n\times n$ Toeplitz matrix:
\beq \label{ToepDet} Z_n(\mathcal{I})= \det \left(T_{p,q}=t_{p-q}\right)_{1\leq p,q\leq n} \eeq
with discrete Fourier coefficients given by:
\bea \label{FourierCoeff} t_0&=&\frac{1}{2\pi}\sum_{j=1}^d(\beta_j-\alpha_j)=\frac{|\mathcal{I}|}{2\pi}\cr
t_k&=&\frac{1}{2\pi}\sum_{j=1}^d e^{ik\frac{\alpha_j+\beta_j}{2}}(\beta_j-\alpha_j)\sin_c\frac{k(\beta_j-\alpha_j)}{2} \,\,\,\,,\,\,\,\,  \forall\, k\neq 0
\eea
where we denoted $\sin_c(x)=\frac{\sin x}{x}$ the cardinal sine function.
\item A real $n$-dimensional integral with logarithmic potential and Vandermonde interactions:
\bea \label{RealInt} Z_n(\mathcal{I})&=&\frac{2^{n(n-1)}}{(2\pi)^n n!}\int_{\mathcal{I}^n} d\theta_1\dots d\theta_n  \prod_{1\leq p<q\leq n}\sin^2\left(\frac{\theta_p-\theta_q}{2}\right)\cr
&=& \frac{2^{n^2}}{(2\pi)^n n!}\int_{\mathcal{J}^n} dt_1\dots dt_n \, \Delta(t_1,\dots,t_n)^2e^{-n\underset{k=1}{\overset{n}{\sum}} \ln(1+t_k^2)}
\eea
where $\Delta(t_1,\dots,t_n)$ is the usual Vandermonde determinant $\Delta(\mathbf{t})=\underset{1\leq p<q\leq n}{\prod}(t_p-t_q)^2$.
\item A Hermitian matrix integral with prescribed eigenvalues support:
\beq Z_n(\mathcal{I})= c_n\int_{\mathcal{N}_n(\mathcal{J})} \frac{dM_n}{\left(\det(I_n+M_n^2)\right)^n}\eeq
where $\mathcal{N}_n$ is the set of Hermitian matrices with eigenvalues belonging to $\mathcal{J}$. The normalizing constant $c_n$ is related to the volume of the unitary group:
\beqq c_n=\frac{1}{(2\pi)^nn!} \frac{1}{\text{Vol } \mathcal{U}_n}=\frac{1}{(2\pi)^nn!}  \frac{(n!)\underset{j=1}{\overset{n-1}{\prod}}j! }{\pi^{\frac{n(n-1)}{2}}}=\frac{1}{2^n\pi^{\frac{n(n+1)}{2}}}\underset{j=1}{\overset{n-1}{\prod}}j!\eeqq
\item A complex $n$-dimensional integral over some segments of the unit circle with Vandermonde interactions:
\beq \label{ComplexInt} Z_n(\mathcal{I})=(-1)^{\frac{n(n+1)}{2}}i^n \int_{\mathcal{T}^n} du_1\dots d u_n\, \Delta(u_1,\dots,u_n)^2 e^{-n\underset{k=1}{\overset{n}{\sum}} \ln u_k}\eeq  
\end{enumerate}
\end{lemma}

\proof{The proof of the previous lemma is rather elementary. The reformulation in terms of Toeplitz determinant is standard \cite{Widom,BS}. Indeed it is well known \cite{BS} that for a function (usually called ``symbol'' in the context of Toeplitz integrals) $f$ measurable on the unit circle we have:
\beaa I_n(f)&=&\frac{1}{(2\pi)^n n!}\int_{\mathcal{I}^n} d\theta_1\dots d\theta_n \left(\prod_{k=1}^n f(e^{i\theta_k})\right)  \prod_{1\leq p<q\leq n}\left|e^{i \theta_p}-e^{i\theta_q}\right|^2\cr
&=&\det \left(T_{p,q}=t_{p-q}\right)_{1\leq p,q\leq n} \text{ with }   t_k=\frac{1}{2\pi}\int_0^{2\pi} f(e^{i\theta})e^{ik\theta}d\theta \,\,,\,\, \forall\, \, k\in\llbracket -(n-1),n-1\rrbracket \cr
\eeaa
Thus, equality between \eqref{IntExp} and \eqref{ToepDet} corresponds to the application of the last identity with $f=\mathds{1}_{\mathcal{T}}$.
Equality between \eqref{IntExp} and \eqref{RealInt} follows from the change of variables $\theta_j=\tan \frac{t_j}{2}$ which is allowed since the support of the angles is included into $(-\pi,\pi)$. With this change of variables we get:
\beqq \left|e^{i \theta_p}-e^{i\theta_q}\right|^2=\frac{4\left( \tan \frac{\theta_p}{2}-\tan \frac{\theta_q}{2}\right)^2}{\left(1+\tan^2\frac{\theta_p}{2}\right)\left(1+\tan^2\frac{\theta_q}{2}\right)}\eeqq
Observing that $d\theta_k$ provides a factor $d\theta_k=\frac{2}{1+t_k^2}dt_k$ and that we have:
\beqq \prod_{1\leq p<q\leq n}\frac{1}{\left(1+t_p^2\right)\left(1+t_q^2\right)}=\prod_{k=1}^n\frac{1}{(1+t_k^2)^{n-1}}\eeqq
immediately gives \eqref{RealInt}. 
Reformulating the real integral \eqref{RealInt} in terms of a Hermitian matrix integral is standard (see \cite{MehtaBook}) from diagonalization $M=U\Lambda U^\dagger$ of normal matrices. We only note here that the support of eigenvalues is prescribed to $\mathcal{J}$. Eventually the volume of the unitary group can be found in \cite{VolumeUnitaryGroup} and equality between \eqref{IntExp} and \eqref{ComplexInt} is straightforward from the change of variables $u_j=e^{i\theta_j}$.}

\medskip

As presented in the previous theorem, the complex \eqref{ComplexInt} and real \eqref{RealInt} integral reformulations of the problem share the important point that the interactions are given by a Vandermonde determinant $\Delta(x)^2$. We stress that this situation is rather exceptional since a (non-affine) change of variables in such integrals does not generally preserve the form of the interactions.

\begin{remark} We inform the reader that part of the results proven in this article have already been presented in \cite{UnitaryMarchal} using the reformulation in terms of the complex integrals \eqref{ComplexInt} in the context of return times for the eigenvalues of a random unitary matrix. Indeed, as one can obviously see from \eqref{RealInt} and \eqref{ComplexInt}, the complex and real integral reformulations share the crucial fact that the interactions between the eigenvalues are of Vandermonde type: $\Delta(\textbf{x})^2$. This implies that correlation functions of boths reformulations satisfy some loop equations and that the topological recursion may be applied to both models. This leads to the fact that the spectral curves found in this article are related by some symplectic transformations to the ones presented in \cite{UnitaryMarchal} and thus that they provide the same sets of free energies (that reconstruct up to some constants $\ln Z_n(\mathcal{I})$). However, we stress that several practical and theoretical issues were disregarded in \cite{UnitaryMarchal} that can be solved using the real integral reformulation:
\begin{enumerate}\item In \cite{UnitaryMarchal}, results from \cite{BorotGuionnetKoz} were used to justify the form of the large $n$ expansions of the correlators and partition functions. However, results from \cite{BorotGuionnetKoz} have only been proved for integrals with support included in $\mathbb{R}$, but not in the case of a generic closed curve in $\mathbb{C}$ as would require the unit circle. Though it is believed that the tools developed in \cite{BorotGuionnetKoz} should remain valid for certain non-real domains of integration, the mathematical proof is still missing.
\item In \cite{UnitaryMarchal}, normalization issues related to the partition function were completely disregarded. Consequently, the large $n$ asymptotic expansions of the Toeplitz determinants  presented in \cite{UnitaryMarchal} lack some constant terms that we provide in this paper. Though the normalization issues were not particularly important for the physics problem studied in \cite{UnitaryMarchal}, they become essential when one wants to compute exact probabilities and compare them with numeric simulations. It turns out that the reformulation \eqref{RealInt} offers, at least in the one interval case, a simple way to deal with the normalization issues by connecting them to a Selberg integral. In the complex setting the connection to a known integral is less obvious and thus the normalization issues were disregarded in \cite{UnitaryMarchal}.
\item Stronger results than those developed for general interactions \cite{BorotGuionnetKoz} regarding the large $n$ expansion and the reconstruction by the topological recursion are available in the case of real integrals with Vandermonde interactions like \eqref{RealInt} in \cite{BG,BG2,BorotGuionnetKoz}. In particular these results allow a proper rigorous mathematical derivation of the full asymptotic expansion of the Toeplitz integrals \eqref{Zn} and provide the explicit expressions of the first orders including the proper normalizing factors.
\item Numerical simulations in \cite{UnitaryMarchal} were only performed to leading order $O(n^2)$ while we are able to match the theoretical results with numeric simulations up to $O\left(n^{-6}\right)$ in this article.  
\end{enumerate}   
\end{remark}

\section{Study of the one interval case \label{Section1Int}}
\subsection{Known results}
When $d=1$, we can use the rotation invariance and take $\beta_1=-\alpha_1=\pi \epsilon$ with $0<\epsilon<1$. For simplicity we shall \textbf{denote in this section $\mathbf{a}\boldsymbol{=}\text{\textbf{tan}} \frac{\boldsymbol{\pi}\boldsymbol{\epsilon}}{\mathbf{2}}$} and a function of $\epsilon$ will equivalently be seen as a function of $a$ and vice-versa depending on the relevance of the parameter in the discussion. In this section, we want to compute the large expansion of the Toeplitz integral:
\beq \label{IntExp1} Z_n(a)\overset{\text{notation}}{\equiv} Z_n(\epsilon)=\frac{1}{(2\pi)^nn!}\int_{[-\pi\epsilon,\pi\epsilon]^n} d\theta_1\dots d\theta_n  \prod_{1\leq p<q\leq n}\left|e^{i \theta_p}-e^{i\theta_q}\right|^2\eeq   
The corresponding Toeplitz determinant reformulation is particularly easy:
\beq \label{ToepDet1} Z_n(\epsilon)=\det \left(T_{p,q}=t_{p-q}\right)_{1\leq p,q\leq n} \eeq
with the discrete Fourier coefficients given by:
\beq \label{FourierCoeff1} t_k=\epsilon\sin_c(k\pi\epsilon) \,\,\,\,,\,\,\,\,  \forall\, k \in \llbracket-(n-1),n-1\rrbracket\eeq
The reformulation in terms of a $n$-fold integral corresponding to the diagonalized form of a Hermitian matrix integral is given by:
\beq \label{RealInt2}Z_n(a)=\frac{2^{n^2}}{(2\pi)^n n!}\int_{[-a,a]^n} dt_1\dots dt_n  \Delta(t_1,\dots,t_n)^2e^{-n\underset{k=1}{\overset{n}{\sum}} \ln(1+t_k^2)}\eeq
Such integrals have been studied by Widom in \cite{Widom} and do not fall in the standard theory of Toeplitz integrals developed by Szeg\"{o}. Indeed the standard case corresponds to a symbol $f$ which is strictly positive and continuous on the unit circle. In that case, the standard theory of Toeplitz determinants can be applied and one would obtain the strong Szeg\"{o} theorem \cite{Szego}:
\beq \frac{1}{n}\ln \det \left(T_n(f)\right)\overset{n\to \infty}{\to}\frac{1}{2\pi}\int_0^{2\pi} \ln(f(e^{i\theta}))d\theta\eeq
When the symbol is discontinuous but remains strictly positive, adaptations of the theory, known as Fisher-Hartwig singularities, have been developed and the convergence of $\frac{1}{n}\ln \det \left(T_n(f)\right)$ is still obtained though formulas get more involved. However, in our present case, the symbol is vanishing on several intervals of the unit circle and the convergence of $\frac{1}{n}\ln \det \left(T_n(f)\right)$ does no longer hold. Indeed, in this case, Widom proved the following proposition \cite{Widom}:

\begin{theorem}[Widom's result] Let $0<\theta_0<\pi$ and define $\mathcal{T}(\theta_0)=\{e^{it},t\in[-\theta_0,\theta_0]\}$. Then we have:
\beaa\ln \det \left(T_n(\mathds{1}_{\mathcal{T}(\theta_0)})\right)&=&n^2\ln\left(\sin\frac{\theta_0}{2}\right)-\frac{1}{4}\ln n-\frac{1}{4}\ln\left(\cos\frac{\theta_0}{2}\right)\cr
&&+3\,\zeta'(-1)+\frac{1}{12}\ln 2+o(1)\eeaa
where $\zeta$ denotes the Riemann zeta function.
\end{theorem} 

We propose in this section to improve Widom's result by providing a mathematical proof of the form of the large $n$ expansion of $\ln Z_n(\epsilon)$ as well as a general way to compute all sub-leading corrections.

\subsection{Spectral curve and limiting eigenvalues density \label{SpecCurveSection}}

Integral \eqref{RealInt2} may be seen as a gas of eigenvalues with Vandermonde interactions and evolving in a potential $V(x)=\ln(1+x^2)$. Consequently, it falls into the category of integrals studied in \cite{BG,BG2,BorotGuionnetKoz}. In particular, the potential is analytic on $\mathbb{R}$ and has only one minimum at $x=0$. Moreover, since the support of the integration is restricted to a compact set $[-a,a]$, the convergence issues are trivial. Under such conditions, it is proved in \cite{Empirical,Empirical2} that the empirical eigenvalues density $\delta_n$ converges almost surely towards an absolutely continuous limiting eigenvalues density:
\beq \delta_n=\frac{1}{n}\sum_{k=1}^n\delta(x-t_k)\underset{n\to \infty}{\to} d\mu_\infty(x)=\nu(x)dx\eeq
whose support is a finite union of intervals in $[-a,a]$. Determining the limiting eigenvalues density can be done in many ways, but since we plan to use the topological recursion, it seems appropriate to derive the limiting eigenvalues density using the loop equations method. We define the following correlation functions:

\begin{definition}[Correlation functions]\label{CorrFunctions} We define the correlation functions by:
\beaa W_{1,a}(x)&=&\left<\sum_{k=1}^n \frac{1}{x-t_k}\right>\cr
W_{p,a}^{n.c.}(x_1,\dots,x_p)&=&\left<\sum_{i_1,\dots,i_p=1}^n \frac{1}{x_1-t_{i_1}}\dots\frac{1}{x_p-t_{i_p}}\right>\cr
W_{p,a}(x_1,\dots,x_p)&=&\left<\sum_{i_1,\dots,i_p=1}^n \frac{1}{x_1-t_{i_1}}\dots\frac{1}{x_p-t_{i_p}}\right>_c
\eeaa
where the average of a function of the eigenvalues is defined by:
\beqq \left<g(t_1,\dots,t_n)\right>=\frac{2^{n^2}}{(2\pi)^n n! Z_n(a)}\int_{[-a,a]^n} dt_1\dots dt_n \, g(t_1,\dots,t_n)\Delta(\mathbf{t})^2e^{-n\underset{k=1}{\overset{n}{\sum}} \ln(1+t_k^2)}\eeqq
The index $\,_c$ stands for ``connected'' or ``cumulant'' in the sense that:
\beaa 
W_{2,a}(x_1,x_2)&=&W_{2,a}^{n.c.}(x_1,x_2)-W_{1,a}(x_1)W_{1,a}(x_2)\cr
W_{3,a}(x_1,x_2,x_3)&=&W_{3,a}^{n.c.}(x_1,x_2,x_3)-W_{1,a}(x_1)W_{2,a}^{n.c.}(x_2,x_3)-W_{1,a}(x_2)W_{2,a}^{n.c.}(x_1,x_3)\cr
&&-W_{1,a}(x_3)W_{2,a}^{n.c.}(x_1,x_2)+2W_{1,a}(x_1)W_{1,a}(x_2)W_{1,a}(x_3)\cr
\normalfont{\text{etc.}}&&
\eeaa
or in a more general way by the inverse relation:
\beqq W_{p,a}^{n.c.}(x_1,\dots,x_p)= \sum_{\mu \,\vdash \{x_1,\dots,x_p\}}\prod_{i=1}^{\text{l}(\mu)}  W_{|\mu_i|,a}(\mu_i)\eeqq
\end{definition}

Integral \eqref{RealInt2} is a Hermitian matrix integral with hard edges at $t=\pm a$. Loop equations for Hermitian matrix integrals with hard edges have been written in many places \cite{HardWall,Hardedges,Hardedges2}. They can also be easily obtained with the integral:
\beq \frac{1}{(2\pi)^n(n!) Z_n(a)}\int_{[-a,a]^n}dt_1\dots d t_n \sum_{j=1}^n\frac{d}{dt_j}\left(\frac{1}{x-t_j} \Delta(t_1,\dots,t_n)^2 e^{-n\underset{k=1}{\overset{n}{\sum}} \ln (1+t_k^2)}\right)\eeq
Indeed, the last integral is equivalent to:
\beq \label{FirstLoop} W_{1,a}^2(x)+W_{2,a}(x,x)-\frac{2nx}{1+x^2}W_{1,a}(x)+n\left<\sum_{k=1}^n\frac{V'(x)-V'(t_k)}{x-t_k}\right>=\frac{c_1(n)}{x-a}-\frac{c_2(n)}{x+a}\eeq
Equation \eqref{FirstLoop} is exact and the coefficients $c_1(n)$ and $c_2(n)$ are given by:
\footnotesize{
\bea c_1(n)&=&\frac{e^{-n\ln(1+a^2)}}{(2\pi)^n(n!) Z_n(a)}\sum_{j=1}^n\int_{[-a,a]^{n-1}} dt_1\dots dt_{j-1} dt_{j+1} \dots d t_n\,\Delta(t_1,\dots,t_{j-1},a,t_{j+1},\dots, t_n)^2 e^{-n\underset{k\neq j}{\overset{n}{\sum}} \ln (1+t_k^2)}\cr
c_2(n)&=&\frac{e^{-n\ln(1+a^2)}}{(2\pi)^n(n!) Z_n(a)}\sum_{j=1}^n\int_{[-a,a]^{n-1}} dt_1\dots dt_{j-1} dt_{j+1} \dots d t_n\,\Delta(t_1,\dots,t_{j-1},-a,t_{j+1},\dots, t_n)^2 e^{-n\underset{k\neq j}{\overset{n}{\sum}} \ln (1+t_k^2)}\cr
&&
\eea}\normalsize{} 
Note that since the integral \eqref{RealInt2} is invariant under the change $\mathbf{t}\to -\mathbf{t}$ we automatically get $c_1(n)=c_2(n)$. In order to obtain the spectral curve of the problem (which is the Stieltjes transform of the limiting eigenvalues density), we take the leading order in $n$ of equation \eqref{FirstLoop}. Results from \cite{BG,BG2,BorotGuionnetKoz} show that $W_{1,a}(x)\underset{n\to \infty}{\sim} nW_{1,a}^{(0)}(x)$ and $W_{2,a}(x_1,x_2)\underset{n\to \infty}{=} O(1)$. Defining \beq y(x)=W_{1,a}^{(0)}(x)-\frac{1}{2}V'(x)=W_{1,a}^{(0)}(x)-\frac{x}{1+x^2}\eeq
we end up with:
\beq \label{LoopProject} y(x)^2=\frac{x^2}{(1+x^2)^2}-\frac{2}{1+x^2}\left(\underset{n\to \infty}{\lim}\left<\frac{1}{n}\sum_{k=1}^n\frac{1}{1+t_k^2}\right>-x\underset{n\to \infty}{\lim}\left<\frac{1}{n}\sum_{k=1}^n\frac{t_k}{1+t_k^2}\right>\right)+\frac{c(a)}{x-a}-\frac{c(a)}{x+a}\eeq
where the constant $c(a)$ is given by $c(a)=\underset{n\to \infty}{\lim}\frac{1}{n}c_1(n)$. Since the integral \eqref{RealInt2} is invariant under $\mathbf{t}\to -\mathbf{t}$ we get that $\left<\underset{k=1}{\overset{n}{\sum}}\frac{t_k}{1+t_k^2}\right>=0$ so that:
\beq \label{LoopProject2} y(x)^2=\frac{x^2}{(1+x^2)^2}-\frac{2d(a)}{1+x^2}+\frac{c(a)}{x-a}-\frac{c(a)}{x+a}\eeq
where $d(a)$ and $c(a)$ are so far undetermined constants (i.e. independent of $x$). Moreover, note that by definition we must have at large $x$: 
\bea W_{1,a}^{(0)}(x)&=&\frac{1}{x}-\underset{n\to \infty}{\lim}\left<\frac{1}{n}\sum_{k=1}^nt_k\right>\frac{1}{x^2}+O\left(\frac{1}{x^3}\right)\cr
&=&\frac{1}{x}+O\left(\frac{1}{x^3}\right) \,\,\, \Rightarrow \,\,\,y(x)=O\left(\frac{1}{x^3}\right)
\eea
since the integral is invariant under $\mathbf{t}\to -\mathbf{t}$. Using the fact that $y^2(x)=O\left(\frac{1}{x^6}\right)$ in \eqref{LoopProject2} provides two independent equations satisfied by $(c(a),d(a))$ that can be explicitly solved. We find:
\beq c(a)=\frac{1}{2a(1+a^2)} \text{  and  } d(a)=\frac{2+a^2}{2(1+a^2)}\eeq
Finally, we get:
\beq y^2(x)=\frac{1+a^2}{(1+x^2)^2(x^2-a^2)}=\frac{1}{\cos^2(\frac{\pi\epsilon}{2})(1+x^2)^2(x^2-\tan^2(\frac{\pi\epsilon}{2}))}\eeq
In other words since $\cos(\frac{\pi\epsilon}{2})>0$ for $\epsilon\in(0,1)$:
\beq \label{SpectralCurve1} y(x)=\frac{1}{\cos(\frac{\pi\epsilon}{2})(1+x^2)\sqrt{x^2-\tan^2(\frac{\pi\epsilon}{2})}}\eeq
This is equivalent to say that the limiting eigenvalues density is given by:
\beq \label{mu}d\mu_\infty(x)= \frac{dx}{\pi\cos(\frac{\pi\epsilon}{2})(1+x^2)\sqrt{\tan^2(\frac{\pi\epsilon}{2})-x^2}}\mathds{1}_{ \left[-\tan \frac{\pi\epsilon}{2},\tan \frac{\pi\epsilon}{2}\right]}(x)\eeq
The last density can be verified numerically using Monte-Carlo simulations:

\begin{center}
\includegraphics[width=8cm]{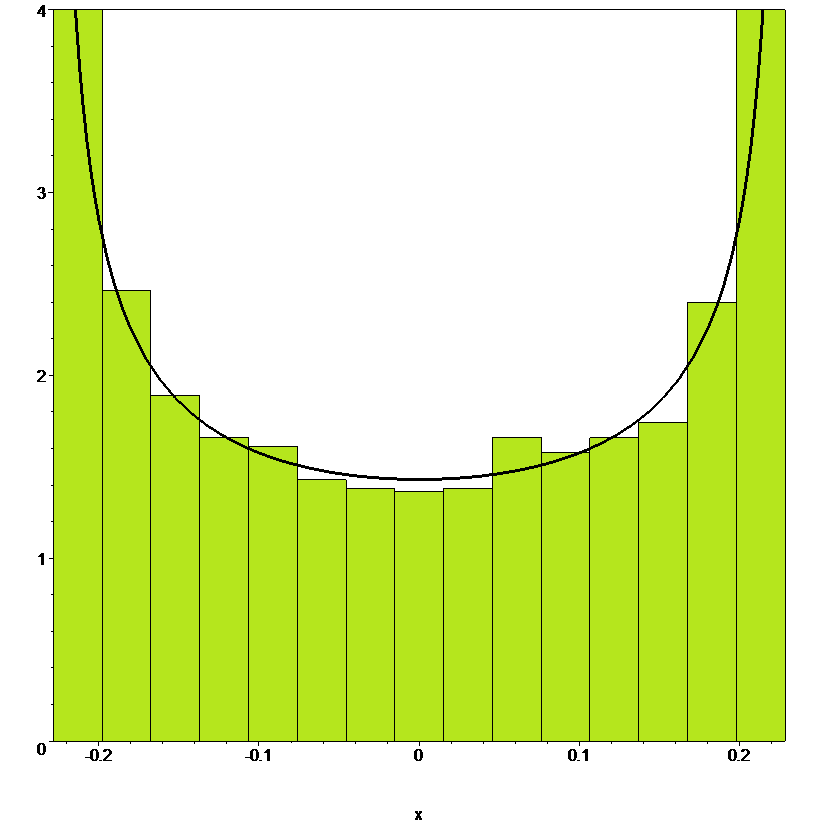}

Fig. $1$: Empirical eigenvalues density obtained from $100$ independent Monte-Carlo simulations of the integral \eqref{RealInt2} in the case $\epsilon=\frac{1}{7}$ and $n=20$. The black curve is the theoretical curve corresponding to \eqref{mu}.
\end{center}

The limiting eigenvalues density is supported on the whole interval $[-\tan\frac{\pi \epsilon}{2},\tan \frac{\pi \epsilon}{2}]$ and exhibits inverse square-root behavior near the edges of the support:
\beq \label{EdgeBehavior}d\mu_\infty(x)\overset{x\to \tan\frac{\pi \epsilon}{2}}{=} O\left(\frac{dx}{\sqrt{x-\tan\frac{\pi \epsilon}{2}}}\right) \text{  and  } d\mu_\infty(x)\overset{x\to -\tan\frac{\pi \epsilon}{2}}{=} O\left(\frac{dx}{\sqrt{x+\tan\frac{\pi \epsilon}{2}}}\right)\eeq
Moreover it is strictly positive inside $[-\tan\frac{\pi \epsilon}{2},\tan \frac{\pi \epsilon}{2}]$. 

\begin{remark} The spectral curve \eqref{SpectralCurve1} is related to the one found in \cite{UnitaryMarchal} (eq. $C.14$) using the complex integral reformulation rather than the real integral reformulation. Indeed, in \cite{UnitaryMarchal}, the spectral curve in the one-interval case is $\td{y}^2=\frac{(\td{x}+1)^2}{4\td{x}^2(\td{x}-e^{i\pi \epsilon})(\td{x}+e^{-i\pi \epsilon})}$. In fact, both curves are equivalent up to the symplectic transformation:
\beq \td{x}=e^{2i\text{Arctan}(x)}\overset{def}{=}f(x) \text{ and } \td{y}=\frac{1}{f'(x)}y=\frac{1+x^2}{2i}e^{-2i\text{Arctan}(x)}y\eeq
This transformation follows from the combination of $x=\tan \frac{\theta}{2}$ and $\td{x}=e^{i\theta}$ corresponding to a simple reparametrization of the $x$-function (which is a trivial case of the symplectic invariance). In particular, as explained in Appendix \ref{AppendixTopRec}, both curves provide the same set of ``symplectic invariants'' $\left(F^{(g)}\right)_{g\geq 0}$ that reconstruct the large $n$ expansion of $\ln Z_n(a)$. However, the Eynard-Orantin differentials attached to both curves differ which is coherent with the fact that the correlation functions are different in both settings.  
\end{remark}

\subsection{General form of the large $n$ expansions}

As soon as the limiting eigenvalue density \eqref{mu} is determined, we may apply the main results of \cite{BG,BG2,BorotGuionnetKoz} to obtain the general form of the large $n$ expansions of the correlators and of the partition function. However, we first need to prove that Hypothesis $1.1$ of \cite{BG} is satisfied so that we may apply the main results of \cite{BG}. We have:

\begin{proposition}\label{PropNew} The following conditions (Hypothesis $1.1$ of \cite{BG}) are met for integral \eqref{RealInt2} (note that in our case, the potential $V$ does not depend on $n$ so that some conditions of \cite{BG} are trivially verified):
\begin{itemize}\item (Regularity): The potential $V$ is continuous on the integration domain $[b_-,b_+]$. 
\item (Confinement of the potential): If $\pm \infty$ belong to the integration domain, then the potential is assumed to be decaying sufficiently fast:
\beqq \liminf_{x\to \pm \infty} \frac{V(x)}{2\ln|x|}>1\eeqq
\item (One cut regime): The support of the limiting eigenvalues density is a single interval $[\alpha_-,\alpha_+]$ not reduced to a point. 
\item (Control of large deviations): The function $x\mapsto \frac{1}{2}V(x)+\int_{\mathbb{R}} |x-\xi|d\mu_\infty(\xi)$ defined on $[b_-,b_+]\setminus(\alpha_-,\alpha_+)$ achieves its minimum only in $\alpha_-$ or $\alpha_+$. 
\item (Off-Criticality): The limiting eigenvalues density is off-critical in the sense that it is strictly positive inside the interior of its support and behaves like $O\left(\frac{1}{\sqrt{x-b_\pm}}\right)$ if $b_\pm$ is a hard edge or like $O\left(\sqrt{x-\alpha_\pm}\right)$ if $\alpha_\pm$ is a soft edge. 
\item (Analyticity): $V$ can be extended to an analytic function inside a neighborhood of $[\alpha_-,\alpha_+]$. 
\end{itemize}
\end{proposition}

\proof{In our case, most of the points required are easily verified:
\begin{itemize}
\item (Regularity): $x\mapsto \ln(1+x^2)$ is obviously continuous on $[b_-,b_+]=[-a,a]$.
\item (Confinement of the potential): No confinement is required since the support is a compact set of $\mathbb{R}$.
\item (One cut regime): This condition directly follows from equation \eqref{mu}.
\item (Control of large deviations): Since $\alpha_-=b_-=-a$ and $\alpha_+=b_+=a$ (the limiting density is supported on the whole integration domain) then $[b_-,b_+]\setminus(\alpha_-,\alpha_+)=\{\alpha_-,\alpha_+\}$ so the condition is trivially realized.
\item (Off-Criticality): We only have two hard edges and equation \eqref{EdgeBehavior} provides the correct behavior. Moreover, we directly observe from its expression that $d\mu_\infty(x)$ is strictly positive inside its support.
\item (Analyticity): $x\mapsto \ln(1+x^2)$ is trivially analytic in a neighborhood of $[-a,a]$.
\end{itemize}
}

\bigskip

Therefore, we can apply the main result of \cite{BG} for $\beta=2$ (as well as Theorems $1.3$ and $1.4$ of \cite{BG2} or results of \cite{BorotGuionnetKoz} for the partition function with hard edges) and we obtain that:

\begin{theorem}[Large $n$ expansions]\label{LargeNExp} The correlators and the partition functions $Z_n(a)$ admit a large $n$ expansion (usually called ``topological expansion'') of the form:
\bea \label{RRR} W_{p,a}(x_1,\dots,x_p)&=&\sum_{g=0}^{\infty} W_{p,a}^{\{2-p-2g\}}(x_1,\dots,x_p)n^{2-p-2g}\cr
 Z_n(a)&=&\frac{n^{n+\frac{1}{4}}}{n!}\,\text{exp}\left(\sum_{k=-2}^\infty \td{F}^{\{k\}}(a)n^{-k} \right)\cr
\ln Z_n(a)&=&-\frac{1}{4}\ln n+\sum_{k=-2}^\infty F^{\{k\}}(a)n^{-k}
\eea
The previous large $n$ asymptotic expansions have the precise meaning that $\forall\, K\geq 0$:
\bea W_{p,a}(x_1,\dots,x_p)&=&\sum_{g=0}^K W_{p,a}^{\{2-p-2g\}}(x_1,\dots,x_p)n^{2-p-2g} + o\left(n^{2-p-2K}\right)\cr
\ln Z_n(a)&=&-\frac{1}{4}\ln n+\sum_{k=-2}^K F^{\{k\}}(a)n^{-k} + o\left(n^{-K}\right) 
\eea
where the $o\left(n^{2-p-2K}\right)$ and $o\left(n^{-K}\right)$ are uniform for $x_1,\dots,x_n$ in any compact set of $[-a,a]$ but are not uniform in $n$ nor $K$. In the rest of the article, series like $\underset{k=-2}{\overset{\infty}{\sum}} f_k n^{-k}$ will always be considered as asymptotic expansions in the sense presented above.
\end{theorem}

Notice that the asymptotic expansion of a given correlation function only involves powers of $n$ with the same parity and that the series expansion of $W_{p,a}$ starts at $\left(\frac{1}{n^{p-2}}\right)$. On the contrary, the large $n$ expansion of the partition function $Z_n(a)$ may involve all powers of $n$ and has an extra factor $n^{n+\frac{1}{4}}$ and $n!$. Indeed, the results of \cite{BG2,BorotGuionnetKoz} providing the r.h.s. of \eqref{RRR} only apply directly to $\frac{(2\pi)^n n!}{2^{n^2}}Z_n(a)$. While, the factors $(2\pi)^n$ and $2^{n^2}$ may be absorbed in the definition of the constants $F^{\{k\}}(a)$, the term $n!$ may not. A direct corollary of the previous theorem is that the coefficients $W_{p,a}^{\{2-p-2g\}}$ of the correlators are obtained from the topological recursion:

\begin{corollary}[Reconstruction of the correlators via the topological recursion]\label{Corrr} For all $p\geq 1$ and $g\geq 0$ we have:
\beq W_{p,a}^{^{\{2-p-2g\}}}(x_1,\dots,x_p)dx_1\dots dx_n=\omega_p^{(g)}(x_1,\dots,x_p)\eeq
where $\omega_p^{(g)}(x_1,\dots,x_p)$ is the $(p,g)$ Eynard-Orantin differential (see Appendix \ref{AppendixTopRec} or \cite{EO}) computed from the application of the topological recursion to the (genus zero) spectral curve \eqref{SpectralCurve1}.
\end{corollary}

Details about the topological recursion are presented for completeness in Appendix \ref{AppendixTopRec} and more can be found in \cite{EO} and \cite{BookEynard}. In particular since the spectral curve \eqref{SpectralCurve1} is of genus zero, it can be parametrized globally on $\overline{\mathbb{C}}=\mathbb{C}\cup\{\infty\}$ via:
\bea x(z)&=&\frac{1}{2}\tan\left(\frac{\pi \epsilon}{2}\right)\left(z+\frac{1}{z}\right)\cr
y(z)&=&\frac{2}{\sin(\frac{\pi \epsilon}{2})\left(1+\frac{1}{4}\tan^2(\frac{\pi \epsilon}{2})\left(z+\frac{1}{z}\right)\right)\left(z-\frac{1}{z}\right)}
\eea
Moreover, the normalized bi-differential $\omega_2^{(0)}(z_1,z_2)$ required to initialize the topological recursion is $\omega_2^{(0)}(z_1,z_2)=\frac{dz_1dz_2}{(z_1-z_2)^2}$ for genus zero curves. The proof of corollary \ref{Corrr} is standard. Indeed, by construction the correlations functions $W_{p,a}(x_1,\dots,x_p)$ satisfy the loop equations arising in Hermitian matrix models. Moreover, Theorem \ref{LargeNExp} ensures that they have a topological expansion and by definition these correlators may only have singularities at the branchpoints or at the edges. These properties are also satisfied by the Eynard-Orantin differentials (See \cite{EO}) and thus both sets must match since, under these conditions (topological expansion and location of the singularities), the loop equations admit a unique solution.

The situation is more complicated for the partition function $Z_n(a)$. Indeed, as discussed in \cite{BG2} and \cite{BorotGuionnetKoz}, only $\frac{\partial \ln Z_n(a)}{\partial a}$ can be matched with $\frac{\partial \ln \tau(a)}{\partial a}$ where $\ln \tau$ is the free energies generating series associated to the spectral curve \eqref{SpectralCurve1} defined by:
\beq \ln \tau(a)=-\sum_{g=0}^\infty F_{\text{Top. Rec.}}^{(g)}(a)\, n^{2-2g}\eeq

The coefficients $\left(F_{\text{Top. Rec.}}^{(g)}(a)\right)_{g\geq 0}$ are computed from the topological recursion and are commonly called ``symplectic invariants'' or ``free energies'' (See Appendix \ref{AppendixTopRec} or \cite{EO} for the formulas). Note that there is a change of convention regarding the indexes between the indexes of the topological recursion (noted $\,^{(g)}$ and corresponding to $n^{2-2g}$) and the indexes of the asymptotic expansion of $Z_n(a)$ (noted $\,^{\{k\}}$ and corresponding to $n^{-k}$). We keep the notation of \cite{EO} for the topological recursion side to avoid confusion (hence the notation is $F_{\text{Top. Rec.}}^{(0)}$, $F_{\text{Top. Rec.}}^{(1)}$ for powers $n^{2}$, $n^0$ and so on). Thus, the coefficients $\left(F^{\{k\}}(a)\right)_{k\geq -2}$ match with $\left(F_{\text{Top. Rec.}}^{(g)}(a)\right)_{g\geq 0}$ only up to some constants terms (in the sense independent of $a$). For applications in topological string theory and integrable systems, these constants are generally disregarded because normalization issues are mostly irrelevant. But in the context of Toeplitz determinants and probability one needs to find a way to compute them. Otherwise, one may only consider ratio of Toeplitz determinants as studied in \cite{RelativeToeplitz}. 
In our case, we may compute these constants from the knowledge of the asymptotic expansion of $\underset{a_0\to 0}{\lim}\ln Z_n(a_0)$ and using results of \cite{IwakiAllCurves} as we will see below.  

\subsection{Normalization and limit $\epsilon \to 0$}

From \cite{BG2} we have for all $(a,a_0)\in\mathbb{R}_+^*\times \mathbb{R}_+^*$:
\bea \label{NewEq} \ln Z_n(a)-\ln Z_n(a_0)&=&-\sum_{k=-2}^{+\infty} F^{\{k\}}(a)n^{-k}+\sum_{k=-2}^{+\infty} F^{\{k\}}(a_0)n^{-k}\cr
&=&-\sum_{g=0}^{+\infty} F_{\text{Top. Rec.}}^{(g)}(a)\, n^{2-2g}+\sum_{g=0}^{+\infty}  F_{\text{Top. Rec.}}^{(g)}(a_0)\, n^{2-2g}
\eea
Note that when $a$ and $a_0$ are positive numbers, the corresponding potentials satisfy conditions of Proposition \ref{PropNew} (in particular the limiting eigenvalues distributions remain one-cut) thus allowing the use of results of \cite{BG2} that directly gives the previous result. In order to have convergent quantities when $a_0\to 0$, we then isolate a trivial divergent term for $g=0$:
\small{\beq \label{NewEqqq} \ln Z_n(a)=-\sum_{g=0}^{+\infty} F_{\text{Top. Rec.}}^{(g)}(a)\, n^{2-2g}+\left(\ln\left(Z_n(a_0)\right)-n^2\ln a_0\right)+\sum_{g=0}^{+\infty} \left(F_{\text{Top. Rec.}}^{(g)}(a_0)+\delta_{g=0}\ln a_0\right)\, n^{2-2g} \eeq}\normalsize{}
In Appendix \ref{Normalization1Cut} we prove that $\left(\ln\left(Z_n(a_0)\right)-n^2\ln a_0\right)$ has a limit when $a_0\to 0$ which is given by a Selberg integral whose asymptotic expansion can be computed exactly. More precisely, we find:
\bea \label{ExactResults}
\ln Z_n(a_0)-n^2\ln(a_0)&\underset{a_0\to 0}{\to}&4\ln (G(n+1))-\ln (G(2n+1))+2n^2\ln 2-n\ln(2\pi)\cr
&=&-\frac{1}{4}\ln n +3\zeta'(-1)+\frac{1}{12}\ln 2 +\sum_{g=1}^\infty \frac{(4-2^{-2g})B_{2g+2}}{2g(2g+2) n^{2g}}\eea
where $G$ is the $G$-Barnes function, $\zeta$ is the Riemann zeta function and $\left(B_k\right)_{k\geq 0}$ are the Bernoulli numbers. 
Moreover, using invariance of the free energies under symplectic transformations (we only need a trivial symplectic transformation but not the difficult and controversial $(x,y)\leftrightarrow (y,-x)$ exchange) we prove in Appendix \ref{LimitFreeEnergies} that the limits when $a_0\to 0$ of the free energies $ \left(F_{\text{Top. Rec.}}^{(g)}(a_0)+\delta_{g=0}n^2\ln a_0\right)_{g\geq 0}$ are given by the free energies of Legendre's spectral curve $y=\frac{1}{\sqrt{x^2-1}}$ whose values have been recently computed in \cite{IwakiAllCurves} for all $g\geq 0$. This term by term convergence is extended to a large $n$ asymptotic expansion in Appendix \ref{AsymptExpansionLimit} using results of \cite{BG2}. We find:
\beq  \label{ExactResults1}
\sum_{g=0}^{+\infty}\left( F_{\text{Top. Rec.}}^{(g)}(a_0) +\delta_{g=0}\ln a_0 \right) \, n^{2-2g}\overset{a_0\to 0}{\to}n^2\ln 2-\sum_{g=2}^{+\infty}\frac{B_{2g}\left(4-2^{-(2g-2)}\right)}{2g(2g-2)n^{2g-2}} 
 \eeq
Note in particular that we have the simplification (to be understood again in the sense of large $n$ asymptotic expansions):
\footnotesize{\beq \label{ExactResults2}\left(\ln\left(Z_n(a_0)\right)-n^2\ln a_0\right)+\sum_{g=0}^{+\infty} \left(F_{\text{Top. Rec.}}^{(g0)}(a_0)+\delta_{g=0}\ln a_0\right)\, n^{2-2g}
\overset{a_0\to 0}{\to} n^2\ln 2-\frac{1}{4}\ln n +3\zeta'(-1)+\frac{1}{12}\ln 2\eeq}\normalsize{}
Thus, we get from \eqref{NewEq} the following final result:
\beq \label{FinalOneCut} \ln Z_n(a)=-\sum_{g=0}^{+\infty} F_{\text{Top. Rec.}}^{(g)}(a)\, n^{2-2g}+n^2\ln 2-\frac{1}{4}\ln n +3\zeta'(-1)+\frac{1}{12}\ln 2\eeq
Application of the topological recursion to the spectral curve \eqref{SpectralCurve1} is standard and is presented in Appendix \ref{DirectAppliTopRec} with details for the specific computations for $\left( F_{\text{Top. Rec.}}^{(0)}(\epsilon),F_{\text{Top. Rec.}}^{(1)}(\epsilon)\right)$ given in Appendix \ref{AppendixF0F1} (equations \eqref{F0Int} and \eqref{F1Int}). We remind here that $a=\tan \frac{\pi\epsilon}{2}$. We obtain the first orders:
\bea \label{ResultTopRec}F_{\text{Top. Rec.}}^{(0)}(\epsilon)&=&\ln 2-\ln\left(\sin\frac{\pi \epsilon}{2}\right)  \cr
F_{\text{Top. Rec.}}^{(1)}(\epsilon)&=&\frac{1}{4}\ln\left(\cos\left(\frac{\pi \epsilon}{2}\right)\right) \cr
F_{\text{Top. Rec.}}^{(2)}(\epsilon)&=&\frac{1}{64}-\frac{1}{32}\tan^2\left(\frac{\pi \epsilon}{2}\right)\cr
F_{\text{Top. Rec.}}^{(3)}(\epsilon)&=&-\frac{1}{256}-\frac{1}{128}\tan^2\left(\frac{\pi \epsilon}{2}\right)-\frac{5}{128}\tan^4\left(\frac{\pi \epsilon}{2}\right)
\eea
so that equation \eqref{FinalOneCut2} gives:
\bea \label{FinalOneCut2} \ln Z_n(a)&=&n^2\ln\left(\sin\left(\frac{\pi \epsilon}{2}\right)\right)-\frac{1}{4}\ln n -\frac{1}{4}\ln\left(\cos\left(\frac{\pi \epsilon}{2}\right)\right)+3\,\zeta'(-1)+\frac{1}{12}\ln 2\cr
&&+\frac{1}{64n^2}\left(2\tan^2\left(\frac{\pi \epsilon}{2}\right)-1\right)+ \frac{1}{256n^4}\left(1+2\tan^2\left(\frac{\pi \epsilon}{2}\right)+10\tan^4\left(\frac{\pi \epsilon}{2}\right)\right)\cr
&&+O\left(\frac{1}{n^6}\right)
\eea

\subsection{Final result}

Using the rotation invariance of the problem, we may easily generalize the previous result for any interval $[\alpha,\beta]$ and we obtain the following theorem:

\begin{theorem}[Asymptotic expansion of Toeplitz determinants in the one interval case\label{Toeplitz1Cut}] For $(\alpha,\beta)$ such that $0<|\beta-\alpha|<2\pi$, the Toeplitz determinant with symbol $f=\mathds{1}_{\mathcal{T}(\alpha,\beta)}$ where $\mathcal{T}(\alpha,\beta)=\{e^{it}, t\in [\alpha,\beta]\}$ admits a large $n$ asymptotic expansion of the form (with the same meaning as the one given in Theorem \ref{LargeNExp}):
\beaa &&\ln \det T_n(\mathds{1}_{\mathcal{T}(\alpha,\beta)})=n^2\ln\left(\sin\left(\frac{|\beta-\alpha|}{4}\right)\right)-\frac{1}{4}\ln n -\frac{1}{4}\ln\left(\cos\left(\frac{|\beta-\alpha|}{4}\right)\right)\cr 
&&+3\,\zeta'(-1)+\frac{1}{12}\ln 2- \sum_{g=2}^\infty F^{(g)}(a)n^{2-2g}\eeaa
where $a=\tan\left(\frac{|\beta-\alpha|}{4}\right)$ and the coefficients $\left(F^{(g)}(a)\right)_{g\geq 2}$ are the Eynard-Orantin free energies (also called symplectic invariants) associated to the spectral curve 
\beqq y^2(x)=\frac{1}{\cos^2\left(\frac{|\beta-\alpha|}{4}\right)(1+x^2)^2\left(x^2-\tan^2\left(\frac{|\beta-\alpha|}{4}\right)\right)}\eeqq
The previous large $n$ asymptotic expansion has the precise meaning given in Theorem \ref{LargeNExp}.  
In particular, the first orders of the expansion are given by:
\beaa \ln \det T_n(\mathds{1}_{\mathcal{T}(\alpha,\beta)})&=&n^2\ln\left(\sin\left(\frac{|\beta-\alpha|}{4}\right)\right)-\frac{1}{4}\ln n-\frac{1}{4}\ln\left(\cos\left(\frac{|\beta-\alpha|}{4}\right)\right)\cr 
&&+3\,\zeta'(-1)+\frac{1}{12}\ln 2+\frac{1}{64n^2}\left(2\tan^2\left(\frac{|\beta-\alpha|}{4}\right)-1\right)\cr
&&+ \frac{1}{256n^4}\left(1+2\tan^2\left(\frac{|\beta-\alpha|}{4}\right)+10\tan^4\left(\frac{|\beta-\alpha|}{4}\right)\right)+O\left(\frac{1}{n^6}\right)
\eeaa
\end{theorem}

\subsection{Numerical study}
We can efficiently compute the Toeplitz determinants \eqref{ToepDet1} numerically up to $n=35$. This allows to compare the theoretical formula \eqref{FinalOneCut2} with the numeric simulations up to order $o\left(\frac{1}{n^4}\right)$. We obtain the following picture:

\begin{center}
\includegraphics[width=16cm]{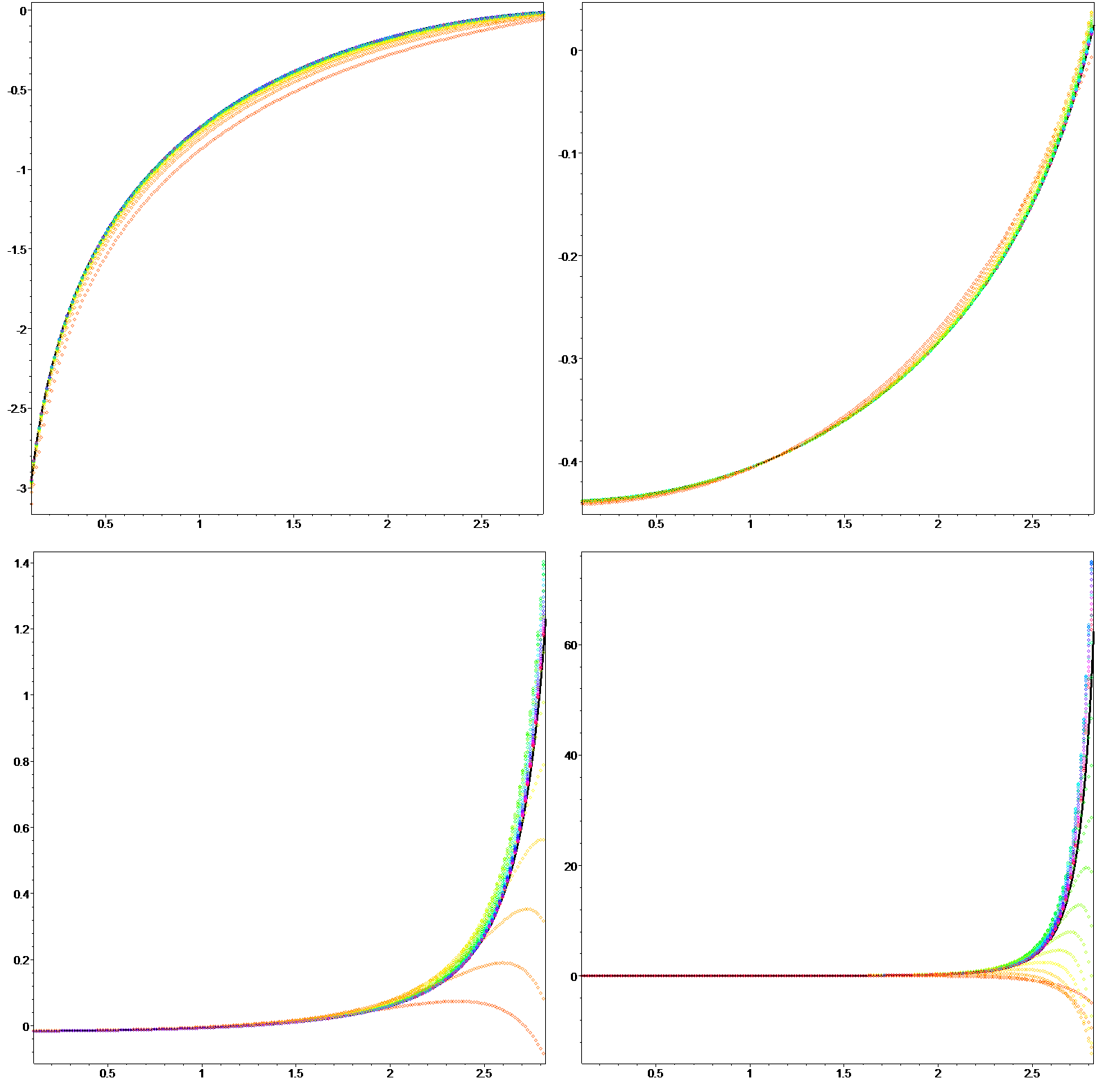}

Fig. $2$: Computations (from \eqref{ToepDet1}) of the Toeplitz determinants $\theta\mapsto \ln Z_n(\frac{\theta}{\pi})$ with $0<\theta<\pi$ for $2\leq n\leq 35$ with subtraction of the first coefficients of the large $n$ expansion \eqref{FinalOneCut2} (Colored dots: starting from orange to yellow, green and purple as $n$ increases). The black curves are the theoretical predictions given by (from top to bottom and from left to right): $\theta\mapsto \ln\left(\sin\left(\frac{\theta}{2}\right)\right)$, $\theta\mapsto -\frac{1}{4}\ln\left(\cos\left(\frac{\theta}{2}\right)\right)+3\,\zeta'(-1)+\frac{1}{12}\ln 2$, $\theta\mapsto \frac{1}{64}\left(2\tan^2\left(\frac{\theta}{2}\right)-1\right)$ and $\theta \mapsto \frac{1}{256}\left(1+2\tan^2\left(\frac{\theta}{2}\right)+10\tan^4\left(\frac{\theta}{2}\right)\right)$. 
\end{center}

We obviously see on the last figure that the numeric simulations are compatible with the theoretical results up to order $o\left(\frac{1}{n^4}\right)$. This provides additional credit for the general formulas proved in Theorem \ref{Toeplitz1Cut} and the reconstruction of the expansion from the topological recursion. Note that the reformulation of $Z_n(a)$ in terms of the determinant of a symmetric Toeplitz matrix is particularly useful since it allows fast computations of $Z_n(a)$ even for relatively large values of $n$. We performed the computations using Maple software and could compute the values of $Z_n(a)$ from $n=2$ to $n=35$ in no more than a few minutes on a standard laptop.

\section{Toeplitz determinants with a discrete rotational symmetry \label{SectionSeveralInt}}

Let $r\geq 0$ be a given integer and let $\epsilon\in (0,1)$ be a given number. In this section we consider the Toeplitz determinants:
\bea \label{ToepR} Z_n(\mathcal{I}_r)&=&\frac{1}{(2\pi)^n n!} \int_{\left(\mathcal{I}_r\right)^n} d\theta_1\dots d\theta_n \prod_{1\leq p<q\leq n}\left|e^{i\theta_p}-e^{i\theta_q}\right|^2 \text{  with  }\cr
\mathcal{I}_r&=&\underset{k=-r}{\overset{r}{\bigcup}}\left[\frac{2\pi k}{2r+1}-\frac{\pi \epsilon}{2r+1}, \frac{2\pi k}{2r+1}+\frac{\pi \epsilon}{2r+1}\right]\eea
For simplicity, we denote $\alpha^{(r)}_k=\frac{2\pi k}{2r+1}-\frac{\pi \epsilon}{2r+1}$, $\beta^{(r)}_k=\frac{2\pi k}{2r+1}+\frac{\pi \epsilon}{2r+1}$ and $\gamma^{(r)}_k=\frac{2\pi k}{2r+1}$ for $-r\leq k\leq r$. We also define:
\beqq a_k^{(r)}=\tan\left(\frac{\alpha_k^{(r)}}{2}\right) \text{  and  } b_k^{(r)}=\tan\left(\frac{\beta_k^{(r)}}{2}\right) \text{  for  } -r\leq k\leq r\eeqq
Note that $Z_n(\mathcal{I}_r)$ can also be interpreted as the probability to obtain all angles $(\theta_j)_{1\leq j\leq n}$ in $\mathcal{I}_r$. Similarly to the last section, we also introduce the sets:
\beaa \mathcal{J}_r&=&\underset{k=-r}{\overset{r}{\bigcup}}\left[a_k^{(r)},b_k^{(r)}\right]=\underset{k=-r}{\overset{r}{\bigcup}} \left[\tan\left( \frac{\pi k}{2r+1}-\frac{\pi \epsilon}{2(2r+1)}\right),\tan\left( \frac{\pi k}{2r+1}+\frac{\pi \epsilon}{2(2r+1)}\right)\right]\cr
\mathcal{T}_r&=& \left\{e^{i\theta},\theta\in \mathcal{I}_r\right\}
\eeaa
Therefore, from Lemma \ref{TheoReform}, $Z_n(\mathcal{I}_r)$ can be reformulated as follow:
\begin{enumerate} \item $Z_n(\mathcal{I}_r)=\det T_n^{(r)}$ with $\left(T_n^{(r)}\right)_{i,j}=t_{i-j}$ the $n\times n$ Toeplitz matrix given by:
\bea \label{Toep2R1} t_0&=&\frac{|\mathcal{I}_r|}{2\pi}=\epsilon\cr
t_k&=& \epsilon \sin_c\left(\frac{\pi \epsilon k}{2r+1}\right) \delta_{k\,\equiv\, 0\,[2r+1]} \,\text{ for }k\neq 0
\eea
Note in particular that the Toeplitz matrices are mostly empty since only bands with indexes multiple of $2r+1$ are non-zero. 
\item A real integral with Vandermonde interactions:
\beq \label{RealRodd} Z_n(\mathcal{I}_r)=\frac{2^{n^2}}{(2\pi)^n n!}\int_{\left(\mathcal{J}_r\right)^n} dt_1\dots dt_n\,  \Delta(t_1,\dots,t_n)^2e^{-n\underset{k=1}{\overset{n}{\sum}} \ln(1+t_k^2)}\eeq
\item A complex integral: 
\beq \label{ComplexRodd} Z_n(\mathcal{I}_r)=(-1)^{\frac{n(n+1)}{2}}i^n \int_{\left(\mathcal{T}_r\right)^n} du_1\dots d u_n\, \Delta(u_1,\dots,u_n)^2 e^{-n\underset{k=1}{\overset{n}{\sum}} \ln u_k}\eeq 
\end{enumerate}

The situation can be illustrated as follow:

\begin{center}
\includegraphics[width=6cm]{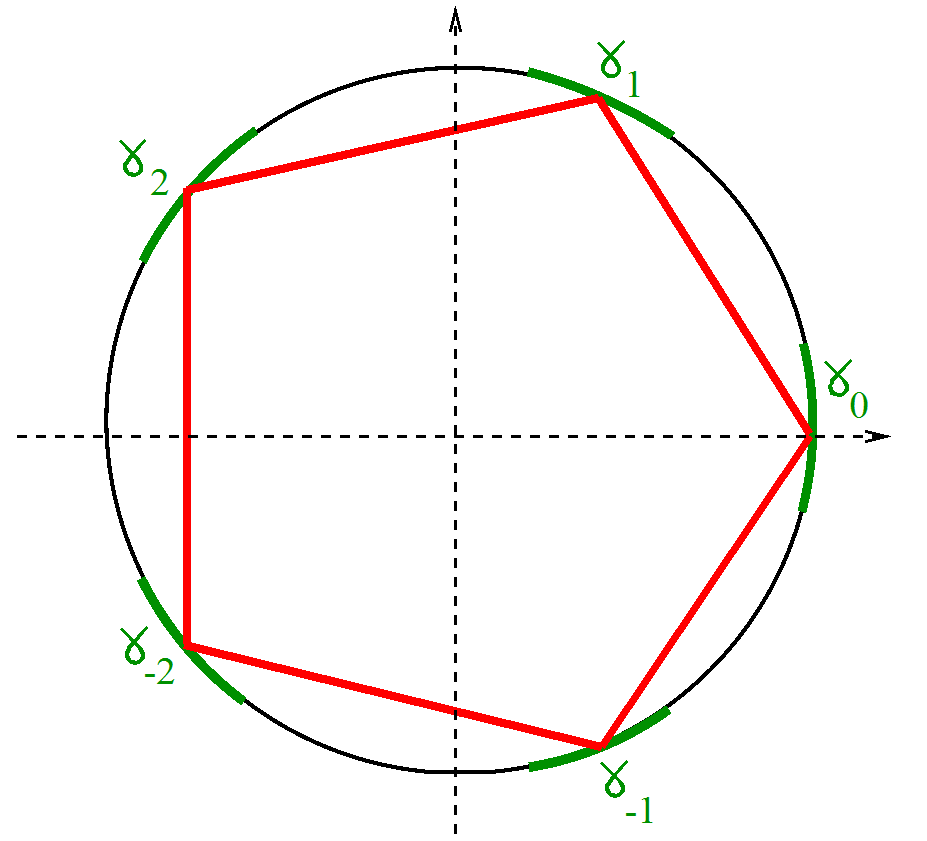}

Fig. $3$: Illustration (in green) of the set $\mathcal{T}_{r=2}$ for $\epsilon=\frac{1}{5}$.
\end{center}

\subsection{Computation of the spectral curve}
We want to compute the spectral curve associated to the integral \eqref{RealRodd}. This integral corresponds to a Hermitian matrix integral with hard edges at $\left(a_k^{(r)}\right)_{-r\leq k\leq r}$  and $\left(b_k^{(r)}\right)_{-r\leq k\leq r}$. Following the same method as in section \ref{SpecCurveSection}, we define $\left<g(\mathbf{t})\right>_r$ as the average of the function $g(\mathbf{t})$ relatively to the measure induced by \eqref{RealRodd}. With the same arguments as in section \ref{SpecCurveSection} , we get a spectral curve of the form:
\bea \label{LoopProjectRodd} y(x)^2&=&\frac{x^2}{(1+x^2)^2}-\frac{2}{1+x^2}\left(\underset{n\to \infty}{\lim}\left<\frac{1}{n}\sum_{j=1}^n\frac{1}{1+t_j^2}\right>_r-x\underset{n\to \infty}{\lim}\left<\frac{1}{n}\sum_{j=1}^n\frac{t_j}{1+t_j^2}\right>_r\right)\cr
&&+\sum_{j=-r}^{r}\left(\frac{A_k}{x-a_k^{(r)}}+\frac{B_k}{x-b_k^{(r)}}\right)\eea
where the constants $A_k$ (resp. $B_k$) are given by $A_k=-\underset{n\to \infty}{\lim}\frac{1}{n}A_{k,n}$ and $B_k=-\underset{n\to \infty}{\lim}\frac{1}{n}B_{k,n}$ with:
\footnotesize{
\bea A_{k,n}&=&-\frac{e^{-n\ln(1+(a_k^{(r)})^2)}}{(2\pi)^n(n!) Z_n(\mathcal{I}_r)}\sum_{j=1}^n\int_{\left(\mathcal{J}_r\right)^{n-1}} \frac{dt_1\dots dt_{j-1} dt_{j+1} \dots d t_n}{a_k^{(r)}-t_j}\Delta(t_1,\dots,t_{j-1},a_k^{(r)},t_{j+1},\dots, t_n)^2 e^{-n\underset{q\neq j}{\overset{n}{\sum}} \ln (1+t_q^2)}\cr
B_{k,n}&=&\frac{e^{-n\ln(1+(b_k^{(r)})^2)}}{(2\pi)^n(n!) Z_n(\mathcal{I}_r)}\sum_{j=1}^n\int_{\left(\mathcal{J}_r\right)^{n-1}} \frac{dt_1\dots dt_{j-1} dt_{j+1} \dots d t_n}{b_k^{(r)}-t_j}\Delta(t_1,\dots,t_{j-1},b_k^{(r)},t_{j+1},\dots, t_n)^2 e^{-n\underset{q\neq j}{\overset{n}{\sum}} \ln (1+t_q^2)}\cr
&&
\eea}\normalsize{}
Note that the integral \eqref{RealRodd} is invariant under $\mathbf{t}\to -\mathbf{t}$, thus we get that $\left<\underset{j=1}{\overset{n}{\sum}}\frac{t_j}{1+t_j^2}\right>_r=0$. Therefore we end up with:
\beq \label{LoopProject2Rodd} y(x)^2=\frac{x^2}{(1+x^2)^2}-\frac{2d}{1+x^2}+\sum_{j=-r}^{r}\left(\frac{A_k}{x-a_k^{(r)}}+\frac{B_k}{x-b_k^{(r)}}\right)\eeq
where the coefficients $d$ and $(A_k,B_k)_{-(r-1)\leq k\leq r-1}$ are so far undetermined and independent of $x$. Observe now that the choice of $\mathcal{I}_r$ implies that we have:
\beqq a_{-k}^{(r)}=-b_k^{(r)} \text{ for all } 0\leq k\leq r\eeqq
Using the invariance of the integral relatively to $\mathbf{t}\mapsto -\mathbf{t}$, we obtain that:
\beq A_{-k}=-B_k \text{ and } B_{-k}=-A_k \text{ for all } 0\leq k\leq r\eeq
Consequently we can reduce the spectral curve to:
\beq \label{LoopProject2RoddReduced} y(x)^2=\frac{x^2}{(1+x^2)^2}-\frac{2d}{1+x^2}+ \sum_{j=-r}^{r}\frac{2b_k^{(r)}B_k}{x^2-(b_k^{(r)})^2}\eeq
Similarly to the one interval case, the study of large $x$ implies that $y^2(x)=O\left(\frac{1}{x^6}\right)$ giving some algebraic relations between the $2r+2$ undetermined coefficients $(d,B_{-r},\dots,B_{r})$. However unlike the one interval case, when $r\geq 1$, these relations are not sufficient to determine completely the coefficients. Indeed, when $r\geq 1$, the number of unknown coefficients is strictly larger than the number of relations obtained from the fact that $y^2(x)=O\left(\frac{1}{x^6}\right)$. Therefore, some additional relations are required. In order to obtain them we perform the following change of variables (denoting $\mathbf{1}=(1,\dots,1)^t \in \mathbb{R}^n$, $\mathbf{t}=(t_1,\dots,t_n)^t$ and $\td{\mathbf{t}}=(\td{t}_1,\dots,\td{t}_n)^t$) in the integral \eqref{RealRodd}. Let $-r\leq j_0\leq r$ :
\bea \label{ChangeVar} \mathbf{t}&=&\tan\left(\text{Arctan}(\td{\mathbf{t}})+\frac{\pi j_0}{2r+1}\mathbf{1}\right)=\frac{\td{\mathbf{t}}+\tan\left(\frac{\pi j_0}{2r+1}\right)}{1-\td{\mathbf{t}} \tan\left(\frac{\pi j_0}{2r+1}\right)}\,\,\Leftrightarrow\cr
\td{\mathbf{t}}&=&\tan\left(\text{Arctan}(\mathbf{t})-\frac{\pi j_0}{2r+1}\mathbf{1}\right)=\frac{\mathbf{t}-\tan\left(\frac{\pi j_0}{2r+1}\right)}{1+\mathbf{t} \tan\left(\frac{\pi j_0}{2r+1}\right)}\eea 
Note that the domain of integration $\mathcal{J}_r$ is invariant under the former change of variables since any interval $\left[a_k^{(r)},b_k^{(r)}\right]$ is mapped to $\left[a_{k-j_0}^{(r)},b_{k-j_0}^{(r)}\right]$ where the indexes $k-j_0$ are to be understood modulo $2r+1$. Then, straightforward computations show that:
\bea dt_p&=&\frac{1+\tan^2\left(\frac{\pi j_0}{2r+1}\right)}{\left(1-\td{t}_p \tan\left(\frac{\pi j_0}{2r+1}\right)\right)^2}d\td{t}_p\cr
1+t_p^2&=&\frac{\left(1+\tan^2\left(\frac{\pi j_0}{2r+1}\right)\right)(1+\td{t}_p^2)}{\left(1-\td{t}_p \tan\left(\frac{\pi j_0}{2r+1}\right)\right)^2}\cr
(t_p-t_q)^2&=&\frac{\left(1+\tan^2\left(\frac{\pi j_0}{2r+1}\right)\right)^2(\td{t}_p-\td{t}_q)^2}{\left(1-\td{t}_p \tan\left(\frac{\pi j_0}{2r+1}\right)\right)^2\left(1-\td{t}_q \tan\left(\frac{\pi j_0}{2r+1}\right)\right)^2}
\eea
Therefore, the general form of the integral remains invariant:
\beaa  Z_n(\mathcal{I}_r)&=&\frac{2^{n^2}}{(2\pi)^n n!}\int_{\left(\mathcal{J}_r\right)^n} dt_1\dots dt_n  \Delta(t_1,\dots,t_n)^2e^{-n\underset{k=1}{\overset{n}{\sum}} \ln(1+t_k^2)}\cr
&=&\frac{2^{n^2}}{(2\pi)^n n!}\int_{\left(\mathcal{J}_r\right)^n} d\td{t}_1\dots d\td{t}_n  \Delta(\td{t}_1,\dots,\td{t}_n)^2e^{-n\underset{k=1}{\overset{n}{\sum}} \ln(1+\td{t}_k^2)}
\eeaa
and thus the function $W_1(x)$ is given by:
\bea \label{EQQ} &&W_1(x)=\frac{2^{n^2}}{(2\pi)^n (n!)Z_n(\mathcal{I}_r)}\int_{\left(\mathcal{J}_r\right)^n} d\mathbf{t}\left(\sum_{k=1}^n\frac{1}{x-t_k}\right)  \Delta(\mathbf{t})^2e^{-n\underset{k=1}{\overset{n}{\sum}} \ln(1+t_k^2)}=\frac{2^{n^2}}{(2\pi)^n (n!)Z_n(\mathcal{I}_r)}\cr
&&\int_{\left(\mathcal{J}_r\right)^n} d\td{\mathbf{t}}\left(\sum_{k=1}^n\frac{1-\td{t}_k\tan\left(\frac{\pi j_0}{2r+1}\right)}{x-\tan\left(\frac{\pi j_0}{2r+1}\right)-\td{t}_k\left(1+x\tan\left(\frac{\pi j_0}{2r+1}\right)\right)}\right)  \Delta(\td{\mathbf{t}})^2e^{-n\underset{k=1}{\overset{n}{\sum}} \ln(1+\td{t}_k^2)}
\eea
We now observe that the equation:
\beq \frac{1-at}{x-a-t(1+ax)}=\mu+\frac{\nu}{x-t}+\frac{\rho\, t}{x^2-t^2}\eeq
where $(\mu,\nu,\rho)$ are independent of $t$ admits the following solution:
\beq (\mu,\nu,\rho,a)=\left(\frac{a}{1+ax},\frac{x(1+a^2)}{(x-a)(1+ax)},-\frac{(1+a^2)(1+x^2)}{(x-a)(1+ax)^2},-\frac{2x}{x^2-1}\right)\eeq
In particular we always have $\frac{\mu}{1-\nu}=\frac{x}{1+x^2}=\frac{1}{2}V'(x)$. Inserting these results with $a=\tan\left(\frac{\pi j_0}{2r+1}\right)$, $t=\td{t}_k$ and:
\bea \left\{
\begin{array}{ccc}
  x_{j_0} &=&\tan\left(\frac{\pi i_0}{2r+1}-\frac{\pi}{2(2r+1)}\right) \,\text{  if  } \,j_0=2i_0 \text{  is even}  \\
  x_{j_0} &=&\tan\left(\frac{\pi i_0}{2r+1}+\frac{\pi}{2(2r+1)}\right) \,\text{  if  } \,j_0=2i_0+1 \text{  is odd}
\end{array}
\right.
\eea
that are solutions of $a(x^2-1)+2x=0$ located outside of $\mathcal{J}_r$, we end up in \eqref{EQQ} with:
\beq W_1(x_{j_0})=\mu+\nu W_1(x_{j_0}) +\rho \left<\underset{k=1}{\overset{n}{\sum}}\frac{t}{x_{j_0}^2-t^2}\right>_r \eeq
Using the fact that $\left<\underset{k=1}{\overset{n}{\sum}}\frac{t}{x^2-t^2}\right>_r=0$ from the symmetry $\mathbf{t}\to-\mathbf{t}$, we get to:
\beq \label{Obs} W_1(x_{j_0})=\mu+\nu W_1(x_{j_0}) \,\,\,\Rightarrow \,\,\, W_1(x_{j_0})=\frac{\mu}{1-\nu}=\frac{x_{j_0}}{1+x_{j_0}^2} \,\,\, \Rightarrow y(x_{j_0})=0\eeq
Hence the values $\left(\tan\left(\frac{\pi j}{2r+1}+\frac{\pi}{2(2r+1)}\right)\right)_{-r\leq j\leq r-1}$ (that are identical to $\left(\tan\left(\frac{\pi j}{2r+1}-\frac{\pi}{2(2r+1)}\right)\right)_{-(r-1)\leq j\leq r}$)) are $2r$ distinct zeros of the function $x\mapsto y(x)$ located outside $\mathcal{J}_r$. This provides $2r$ distinct double zeros for $y^2(x)$. From \eqref{LoopProject2RoddReduced} and the fact that $y^2(x)\underset{x\to \infty}{=}O\left(\frac{1}{x^6}\right)$ we get that the spectral curve must be of the form:
\beq y^2(x)=\frac{P_{4r+4}(x)}{(1+x^2)^2\underset{k=-r}{\overset{r}{\prod}}(x^2-(b_k^{(r)})^2)} \text{ with } P_{4r+4} \text{ a polynomial of degree } 4r\eeq
Since we have found $4r$ zeros (counted with their multiplicities), we have:
\beq y^2(x)=\lambda_r \frac{\underset{k=0}{\overset{r-1}{\prod}} \left(x^2-\tan^2\left(\frac{\pi k}{2r+1}+\frac{\pi}{2(2r+1)}\right)\right)^2} {(1+x^2)^2\underset{k=-r}{\overset{r}{\prod}}\left(x^2-\tan^2\left(\frac{\pi k}{2r+1}+\frac{\pi \epsilon}{2(2r+1)}\right)\right)}\eeq
with the convention that for $r=0$ we take empty products (like $\underset{k=0}{\overset{-1}{\prod}}\left(x^2-\tan^2\left(\frac{\pi k}{2r+1}+\frac{\pi}{2(2r+1)}\right)\right)^2$) equal to $1$. The constant $\lambda_r$ can be determined using the behavior of $y^2(x)$ around $\pm i$. Indeed, by definition the function $z\mapsto W_1(z)$ is analytic in a neighborhood of $\pm i$. Consequently, the poles of $y^2(x)$ at $x=\pm i$ only comes from the shift by $-\frac{1}{2}V'(x)=\frac{x}{1+x^2}$. Therefore we should get $y^2(x)\underset{x\to i}{\sim} \frac{1}{4(x-i)^2}$. We get:
\beq \frac{1}{4}=\lambda_r\frac{\underset{k=0}{\overset{r-1}{\prod}} \left(1+\tan^2\left(\frac{\pi k}{2r+1}+\frac{\pi}{2(2r+1)}\right)\right)^2} {4\underset{k=-r}{\overset{r}{\prod}}\left(1+\tan^2\left(\frac{\pi k}{2r+1}+\frac{\pi \epsilon}{2(2r+1)}\right)\right)} \,\Leftrightarrow\,
 \lambda_r=\frac{\underset{k=0}{\overset{r-1}{\prod}} \cos^4\left(\frac{\pi k}{2r+1}+\frac{\pi}{2(2r+1)}\right)} {\underset{k=-r}{\overset{r}{\prod}}\cos^2\left(\frac{\pi k}{2r+1}+\frac{\pi \epsilon}{2(2r+1)}\right)}\eeq
Finally we get:
\bea \label{SpecCurve2R1} y^2(x)&=&\frac{\underset{k=0}{\overset{r-1}{\prod}} \cos^4\left(\frac{\pi k}{2r+1}+\frac{\pi}{2(2r+1)}\right)} {\underset{k=-r}{\overset{r}{\prod}}\cos^2\left(\frac{\pi k}{2r+1}+\frac{\pi \epsilon}{2(2r+1)}\right)} \frac{\underset{k=0}{\overset{r-1}{\prod}} \left(x^2-\tan^2\left(\frac{\pi k}{2r+1}+\frac{\pi}{2(2r+1)}\right)\right)^2} {(1+x^2)^2\underset{k=-r}{\overset{r}{\prod}}\left(x^2-\tan^2\left(\frac{\pi k}{2r+1}+\frac{\pi \epsilon}{2(2r+1)}\right)\right)}\cr
&=&\frac{\underset{k=0}{\overset{r-1}{\prod}} \left(x^2\cos^2\left(\frac{\pi k}{2r+1}+\frac{\pi}{2(2r+1)}\right)-\sin^2\left(\frac{\pi k}{2r+1}+\frac{\pi}{2(2r+1)}\right)\right)^2} {(1+x^2)^2\underset{k=-r}{\overset{r}{\prod}}\left(x^2\cos^2\left(\frac{\pi k}{2r+1}+\frac{\pi \epsilon}{2(2r+1)}\right)-\sin^2\left(\frac{\pi k}{2r+1}+\frac{\pi \epsilon}{2(2r+1)}\right)\right)}\cr
&=&\frac{\underset{k=0}{\overset{r-1}{\prod}} \left((x^2+1)\cos^2\left(\frac{\pi k}{2r+1}+\frac{\pi}{2(2r+1)}\right)-1\right)^2} {(1+x^2)^2\underset{k=-r}{\overset{r}{\prod}}\left((x^2+1)\cos^2\left(\frac{\pi k}{2r+1}+\frac{\pi \epsilon}{2(2r+1)}\right)-1\right)}
\eea
The corresponding limiting eigenvalues density is therefore given by:
\bea \label{mu2R1} d\mu_\infty(x)&=&\frac{\underset{k=0}{\overset{r-1}{\prod}} \cos^2\left(\frac{\pi k}{2r+1}+\frac{\pi}{2(2r+1)}\right)} {\underset{k=-r}{\overset{r}{\prod}}\cos\left(\frac{\pi k}{2r+1}+\frac{\pi \epsilon}{2(2r+1)}\right)} \frac{\underset{k=0}{\overset{r-1}{\prod}} \left|\tan^2\left(\frac{\pi k}{2r+1}+\frac{\pi}{2(2r+1)}\right)-x^2\right|} {\pi(1+x^2)\underset{k=-r}{\overset{r}{\prod}}\sqrt{\left| \tan^2\left(\frac{\pi k}{2r+1}+\frac{\pi \epsilon}{2(2r+1)}\right)-x^2\right|}} \mathds{1}_{\mathcal{J}_r}(x)dx\cr
&=&\frac{\underset{k=0}{\overset{r-1}{\prod}} \left|(1+x^2)\cos^2\left(\frac{\pi k}{2r+1}+\frac{\pi}{2(2r+1)}\right)-1\right|} {\pi(1+x^2)\underset{k=-r}{\overset{r}{\prod}}\sqrt{\left|(1+x^2)\cos^2\left(\frac{\pi k}{2r+1}+\frac{\pi \epsilon}{2(2r+1)}\right)-1\right|}}\mathds{1}_{\mathcal{J}_r}(x)dx \eea
In particular, one can verify that it is properly normalized: $\int_{\mathcal{J}_r} d\mu_\infty(x)=1$. Note that in the case $r=0$, we recover the spectral curve for a single interval \eqref{SpectralCurve1}. We can verify numerically the last limiting eigenvalues density with Monte-Carlo simulations:

\begin{center}
\includegraphics[width=12cm]{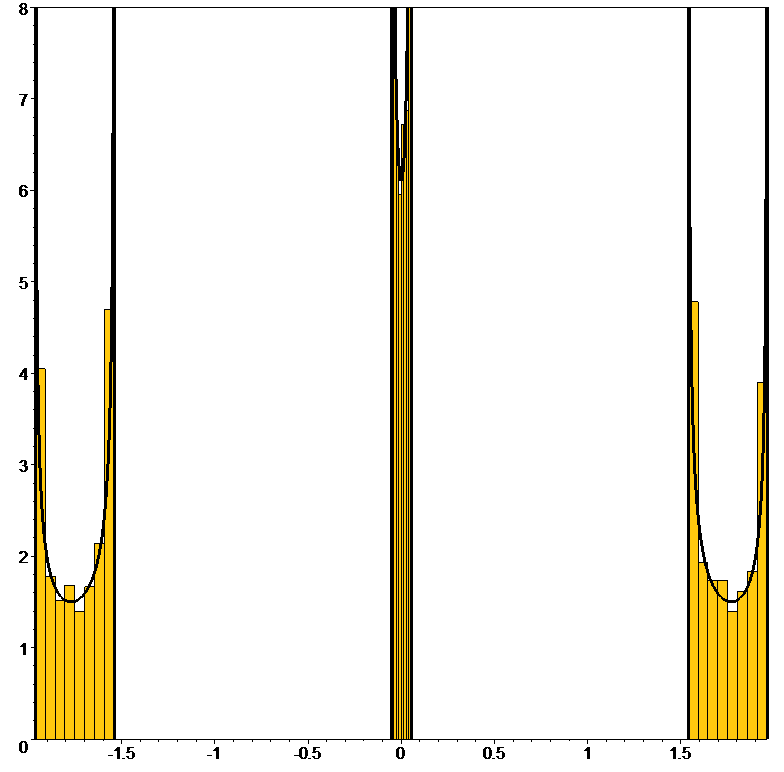}

Fig. $4$: Histogram of $50$ independent simulations of the eigenvalues density induced by \eqref{RealRodd} in the case $r=1$, $\epsilon=\frac{1}{10}$ and $n=60$. The black curve is the theoretical limiting eigenvalues density computed in equation \eqref{mu2R1} 
\end{center}

For a clearer view, it is also interesting to zoom on each interval:

\begin{center}
\includegraphics[width=16cm]{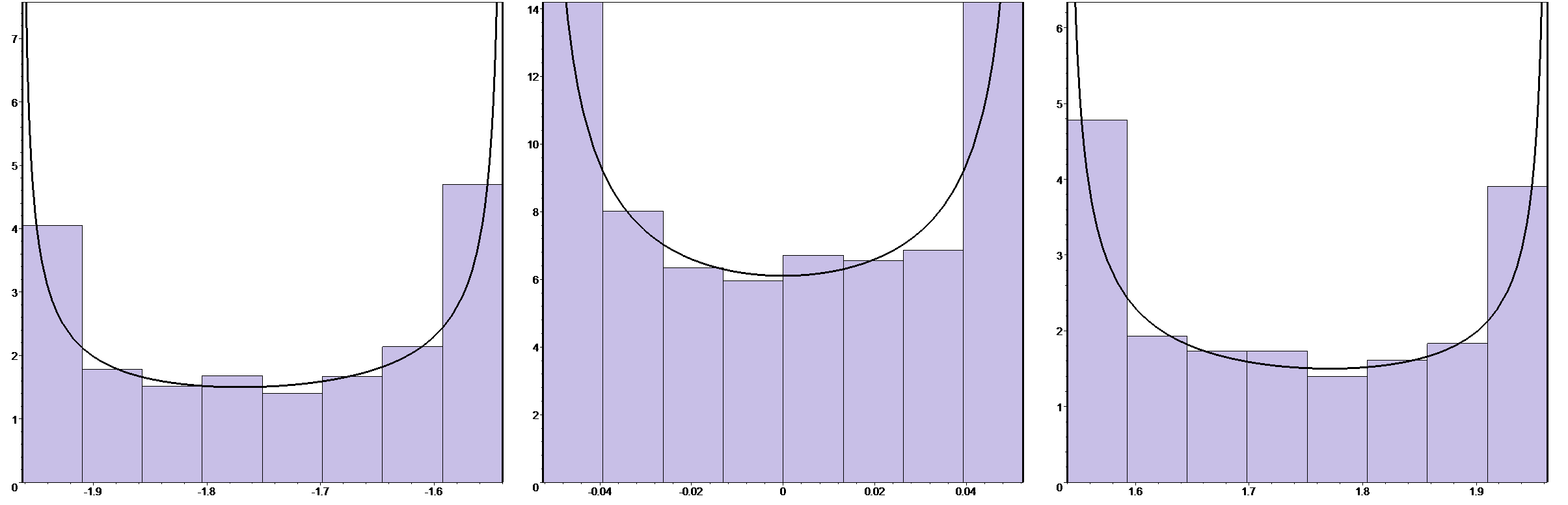}

Fig. $5$: Zoom of the previous histogram in each of the $3$ intervals. The black curve is the theoretical limiting eigenvalues density computed in equation \eqref{mu2R1}.
\end{center}

\subsection{Filling fractions\label{SectionFillingFraction2R1}}
We want to determine the proportion of eigenvalues in each of the $2r+1$ intervals of the limiting eigenvalues density. We start with the previous result:
\beq \label{SpecCurve2R1bis}ydx=\lambda_r \frac{\underset{j=-r}{\overset{r-1}{\prod}} \left(x-\tan\left(\frac{j\pi}{2r+1}+\frac{\pi}{2(2r+1)}\right)\right)}{(1+x^2)\underset{k=-r}{\overset{r}{\prod}} \sqrt{\left(x-\tan\left(\frac{\pi k}{2r+1}-\frac{\pi \epsilon}{2(2r+1)}\right)\right)\left(x-\tan\left(\frac{\pi k}{2r+1}+\frac{\pi \epsilon}{2(2r+1)}\right)\right)}}dx\eeq
and observe that the filling fraction corresponding to the $k_0^{\text{th}}$ interval, with $-r\leq k_0\leq r$, is given by:
\beq \boldsymbol{\epsilon}_{k_0+r+1}\overset{\text{def}}{=}\frac{1}{i\pi}\int_{\tan\left(\frac{\pi k_0}{2r+1}-\frac{\pi \epsilon}{2(2r+1)}\right)}^{\tan\left(\frac{\pi k_0}{2r+1}+\frac{\pi \epsilon}{2(2r+1)}\right)} d\mu_\infty(x)\eeq
Let $-r\leq j_0\leq r$ be a given integer with $j_0\neq 0$. We perform the change of variables $x=\frac{\td{x}-\tan\frac{\pi j_0}{2r+1}}{1+\td{x}\tan \frac{\pi j_0}{2r+1}}$ described above in the previous integral. Observe in particular that we have $\frac{dx}{1+x^2}=\frac{d\td{x}}{1+\td{x}^2}$. Moreover under this change of variable, the interval $x\in \left[\tan\left(\frac{\pi k}{2r+1}-\frac{\pi \epsilon}{2(2r+1)}\right), \tan\left(\frac{\pi k}{2r+1}+\frac{\pi \epsilon}{2(2r+1)}\right)\right]$ is directly mapped to the interval $\td{x}\in\left[\tan\left(\frac{\pi (k+j_0)}{2r+1}-\frac{\pi \epsilon}{2(2r+1)}\right), \tan\left(\frac{\pi (k+j_0)}{2r+1}+\frac{\pi \epsilon}{2(2r+1)}\right)\right]$. We also have the identities:
\footnotesize{\bea \left(x-\tan\left(\frac{\pi k}{2r+1}\pm\frac{\pi \epsilon}{2(2r+1)}\right)\right)&=& \frac{\left(\td{x}-\tan\left(\frac{\pi (k+j_0)}{2r+1}\pm\frac{\pi \epsilon}{2(2r+1)}\right)\right) \left(1- \tan\left(\frac{\pi k}{2r+1}\pm\frac{\pi \epsilon}{2(2r+1)}\right)\tan\left(\frac{\pi j_0}{2r+1}\right)\right)}{1+\td{x}\tan\left(\frac{\pi j_0}{2r+1}\right)}\cr
\left(x-\tan\left(\frac{i_0\pi}{2r+1}+\frac{\pi}{2(2r+1)}\right)\right)&=&\frac{\left(\td{x}-\tan\left(\frac{(i_0+j_0)\pi}{2r+1}+\frac{\pi}{2(2r+1)}\right)\right)\left(1-\tan\left(\frac{\pi i_0}{2r+1}+\frac{\pi }{2(2r+1)}\right)\tan\left(\frac{\pi j_0}{2r+1}\right)\right)}{1+\td{x}\tan\left(\frac{\pi j_0}{2r+1}\right)} \cr
&&\text{ if } i_0+j_0\not\equiv r[2r+1]\cr
&&\eea}\normalsize{}
When $i_0+j_0\equiv r[2r+1]$ then we get instead:
\bea\left(x-\tan\left(\frac{i_0\pi}{2r+1}+\frac{\pi}{2(2r+1)}\right)\right)&=& \frac{-\tan\left(\frac{\pi j_0}{2r+1}\right)-\tan\left(\frac{i_0\pi}{2r+1}+\frac{\pi}{2(2r+1)}\right)}{1+\td{x}\tan\left(\frac{\pi j_0}{2r+1}\right)}\cr
&=&\frac{-\tan\left(\frac{\pi j_0}{2r+1}\right)-\frac{1}{\tan\left(\frac{\pi j_0}{2r+1}\right)}}{1+\td{x}\tan\left(\frac{\pi j_0}{2r+1}\right)}\cr
&=&\frac{-1}{\cos\left(\frac{\pi j_0}{2r+1}\right)\sin\left(\frac{\pi j_0}{2r+1}\right)\left(1+\td{x}\tan\left(\frac{\pi j_0}{2r+1}\right)\right)} 
\eea
Observe now that the powers of $1+\td{x}\tan\left(\frac{\pi j_0}{2r+1}\right)$ produce a factor with power $1$ at the numerator and that we can express it like:
\beq 1+\td{x}\tan\left(\frac{\pi j_0}{2r+1}\right)=\tan\left(\frac{\pi j_0}{2r+1}\right)\left(\td{x}-\tan\left(\frac{(r+j_0)\pi}{2r+1}+\frac{\pi}{2(2r+1)}\right)\right)\eeq
Therefore collecting the terms depending on $\td{x}$ we get:
\small{\bea &&\prod_{k=-r}^{r}\sqrt{\left(\td{x}-\tan\left(\frac{\pi (k+j_0)}{2r+1}-\frac{\pi \epsilon}{2(2r+1)}\right)\right)\left(\td{x}-\tan\left(\frac{\pi (k+j_0)}{2r+1}+\frac{\pi \epsilon}{2(2r+1)}\right)\right)}\cr
&&=\prod_{k=-r}^{r}\sqrt{\left(\td{x}-\tan\left(\frac{\pi k}{2r+1}-\frac{\pi \epsilon}{2(2r+1)}\right)\right)\left(\td{x}-\tan\left(\frac{\pi k}{2r+1}+\frac{\pi \epsilon}{2(2r+1)}\right)\right)}\cr
&& \left(\td{x}-\tan\left(\frac{(r+j_0)\pi}{2r+1}+\frac{\pi}{2(2r+1)}\right)\right)\prod_{i_0=-r, i_0+j_0\not\equiv r [2r+1] }^{r-1}\left(\td{x}-\tan\left(\frac{(i_0+j_0)\pi}{2r+1}+\frac{\pi}{2(2r+1)}\right)\right)\cr
&&=\prod_{i_0=-r, i+j_0\not\equiv r [2r+1] }^{r}\left(\td{x}-\tan\left(\frac{(i_0+j_0)\pi}{2r+1}+\frac{\pi}{2(2r+1)}\right)\right)\cr
&&=\underset{i_0=-r}{\overset{r-1}{\prod}} \left(\td{x}-\tan\left(\frac{i_0\pi}{2r+1}+\frac{\pi}{2(2r+1)}\right)\right)
\eea}\normalsize{}
Thus we obtain:
\beq \boldsymbol{\epsilon}_{k_0+r+1}=C_{j_0}\boldsymbol{\epsilon}_{k_0+j_0+r+1}\eeq
where the constant $C_{j_0}$ is given by:
\footnotesize{\bea C_{j_0}&=&-\frac{\underset{i_0=-r, i_0+j_0\not\equiv r [2r+1] }{\overset{r-1}{\prod}} \left(1-\tan\left(\frac{\pi i_0}{2r+1}+\frac{\pi }{2(2r+1)}\right)\tan\left(\frac{\pi j_0}{2r+1}\right)\right)}{\cos^2\left(\frac{\pi j_0}{2r+1}\right) \underset{k=-r}{\overset{r}{\prod}}\sqrt{\left(1- \tan\left(\frac{\pi k}{2r+1}-\frac{\pi \epsilon}{2(2r+1)}\right)\tan\left(\frac{\pi j_0}{2r+1}\right)\right)\left(1- \tan\left(\frac{\pi k}{2r+1}+\frac{\pi \epsilon}{2(2r+1)}\right)\tan\left(\frac{\pi j_0}{2r+1}\right)\right)}}\cr
&=&-\frac{\underset{i_0=-r, i_0+j_0\not\equiv r [2r+1] }{\overset{r-1}{\prod}} \frac{\cos\left(\frac{\pi(i_0+j_0)}{2r+1}+\frac{\pi}{2(2r+1)}\right)}{\cos\left(\frac{\pi i_0}{2r+1}\right)\cos\left(\frac{\pi j_0}{2r+1}\right) }}{\cos^2\left(\frac{\pi j_0}{2r+1}\right) \underset{k=-r}{\overset{r}{\prod}}\sqrt{\frac{\cos\left(\frac{\pi (k+j_0)}{2r+1}+\frac{\pi \epsilon}{2r+1}\right)\cos\left(\frac{\pi (k+j_0)}{2r+1}-\frac{\pi \epsilon}{2r+1}\right)}{\cos\left(\frac{\pi k}{2r+1}-\frac{\pi \epsilon}{2r+1}\right) \cos\left(\frac{\pi k}{2r+1}+\frac{\pi \epsilon}{2r+1}\right)\cos^2\left(\frac{\pi j_0}{2r+1}\right)}}  }=\left(\cos\left(\frac{\pi j_0}{2r+1}\right)\right)^{-2-(2r-1)+(2r+1)}\cr
&=&1
\eea}\normalsize{} 
Hence we have just proved that all filling fractions are equal:

\begin{proposition}\label{FillingFractions} The filling fractions are the same in each interval:
\beqq \forall\,  1\leq k\leq 2r+1\,\,:\,\,\boldsymbol{\epsilon}_k=\frac{1}{2r+1} \eeqq 
\end{proposition}

\begin{remark} This result is compatible with the one found in \cite{UnitaryMarchal} where the filling fractions were also proved to be equal.
\end{remark}

\subsection{General form of the large $n$ asymptotic \label{SectionOddd}}

After determining the limiting eigenvalues density, the next step is to determine the general form of the asymptotic of the correlation functions and of the partition function. However, since the spectral curve is of strictly positive genus ($\genus=2r$), the general form of the asymptotic is more complicated than in the one interval case. Using the main result of \cite{BG2} or \cite{BorotGuionnetKoz} we can prove the following:

\begin{theorem} \label{PropBorot} We have the following large $n$ expansion:
\bea \label{GeneralForm2R1}&&Z_n(\mathcal{I}_r)=\frac{n^{n+\frac{1}{4}(2r+1)}}{n!} \text{exp}\left(\sum_{k=-2}^\infty n^{-k}F_{\boldsymbol{\epsilon}^\star}^{\{k\}}\right)\cr
&&\left\{\sum_{m\geq 0}\,\sum_{\begin{subarray}{l}l_1,\dots,l_m\geq 1\\ k_1,\dots,k_m\geq -2 \\ \underset{i=1}{\overset{m}{\sum}} (l_i+k_i)>0\end{subarray}} \frac{n^{-\left(\underset{i=1}{\overset{m}{\sum}} (l_i+k_i)\right)}}{m!}
\left(\underset{i=1}{\overset{m}{\bigotimes}} \frac{ F^{\{k_i\},(l_i)}_{\boldsymbol{\epsilon}^\star}}{l_i!}\right) \cdot \nabla_\nu^{\otimes\left(\underset{i=1}{\overset{m}{\sum}} l_i\right)} \right\}\Theta_{-n\boldsymbol{\epsilon}^\star}\left(\mathbf{0}\big|F^{\{-2\},(2)}_{\boldsymbol{\epsilon}^\star}\right)\cr
&&\eea
where $\Theta$ is the Siegel theta function:
\beqq \Theta_{\boldsymbol{\gamma}}(\boldsymbol{\nu},\mathbf{T})=\sum_{\mathbf{m}\in \mathbb{Z}^g} \text{exp}\left(-\frac{1}{2}(\mathbf{m}+\boldsymbol{\gamma})^t\,\cdot \mathbf{T}\cdot(\mathbf{m}+\boldsymbol{\gamma})+\boldsymbol{\nu}^t\,\cdot (\mathbf{m}+\boldsymbol{\gamma})\right)\eeqq
and $F^{\{2k\},(l)}_{\boldsymbol{\epsilon}}$ are defined as the $l^{\text{th}}$ derivative of the coefficient $F_{\boldsymbol{\epsilon}}^{\{2k\}}$ relatively to the filling fractions $\boldsymbol{\epsilon}=(\epsilon_1,\dots,\epsilon_{2r+1})^t \in\{\mathbf{u}\in (\mathbb{Q}_+)^{2r+1}\,/ \, \underset{i=1}{\overset{n}{\sum}}u_i=1\}$. The ``optimal'' filling fractions $\boldsymbol{\epsilon}^\star$ corresponds to the vector of filling fractions (i.e. the proportion of eigenvalues in each interval of the support of the limiting eigenvalues density) of the limiting eigenvalues density:
\beqq \boldsymbol{\epsilon}^\star=\left(\frac{1}{2r+1},\dots,\dots,\frac{1}{2r+1}\right)^t\eeqq
Again large $n$ asymptotic expansions presented in this theorem are to be understood as asymptotic expansions up to any arbitrary large negative power of $n$ as in Theorem \ref{LargeNExp}. 
\end{theorem}

The proof of the last theorem consists in verifying the conditions required to apply the main theorem of \cite{BG2} and \cite{BorotGuionnetKoz}. The conditions are very similar to the one-interval case but for completeness we summarize them here:
\begin{itemize}\item (Regularity): The potential $V$ is continuous on the integration domain $\underset{k=-r}{\overset{r}{\bigcup}}[b_-^{(k)},b_+^{(k)}]$. In our case $x\mapsto \ln(1+x^2)$ is obviously continuous on $\mathbb{R}$.
\item (Confinement of the potential): Not required since the integration domain is a compact set of $\mathbb{R}$.
\item (Genus $2r$ regime): The support of the limiting eigenvalues density is given by the union of $2r+1$ single intervals $[\alpha_-^{(k)},\alpha_+^{(k)}]$ not reduced to a point. This is trivial from the explicit expression \eqref{mu2R1}.
\item (Control of large deviations): The function $x\mapsto \frac{1}{2}V(x)+\int_{\mathbb{R}} |x-\xi|d\mu_\infty(\xi)$ defined on $\left(\underset{k=-r}{\overset{r}{\bigcup}}[b_-^{(k)},b_+^{(k)}]\right)\setminus\left(\underset{k=-r}{\overset{r}{\bigcup}}(\alpha_-^{(k)},\alpha_+^{(k)})\right)$ achieves its minimum only at the endpoints $\alpha_-^{(k)}$ or $\alpha_+^{(k)}$. In our case this condition is trivial since the limiting eigenvalues density spans the whole integration domain.
\item (Off-Criticality): The limiting eigenvalues density is off-critical in the sense that it is strictly positive inside the interior of its support and behaves like $O\left(\frac{1}{\sqrt{x-b_\pm}}\right)$ if $b_\pm$ is a hard edge or like $O\left(\sqrt{x-\alpha_\pm}\right)$ if $\alpha_\pm$ is a soft edge. In our case, we have $2(2r+1)$ hard edges and the explicit expression \eqref{mu2R1} provides the correct behavior. Moreover it is obvious from \eqref{mu2R1} that $d\mu_\infty(x)$ is strictly positive inside its support.
\item (Analyticity): $V$ can be extended into an analytic function inside a neighborhood of the integration domain. In our case, $x\mapsto \ln(1+x^2)$ is obviously analytic in a neighborhood of any compact set of $\mathbb{R}$.
\end{itemize}

As one can see, the general form of the large $n$ expansion in the multi-cut regime is much more complicated than in the one-cut regime since eigenvalues may move from an interval to another and thus a summation on the filling fractions is necessary and adds new terms (given by the Siegel Theta function). 
In particular, orders $O(1)$ and beyond exhibit a far more complex structure than in the one-cut case. Indeed, equation \ref{GeneralForm2R1} and the fact that the present situation has a discrete symmetry (in particular $\boldsymbol{\epsilon}^\star=\left(\frac{1}{2r+1},\dots,\frac{1}{2r+1}\right)^t$) imply that these orders depend on the remainder of the Euclidean division of $n$ by $2r+1$.

\medskip 

The term $\text{exp}\left(\underset{k=-1}{\overset{\infty}{\sum}} n^{-2k}F_{\boldsymbol{\epsilon}^\star}^{\{2k\}}\right)$ is usually called the ``perturbative'' or ``formal'' part of the expansion. It is connected to the symplectic invariants computed from the topological recursion by the following proposition (See \cite{BG2} for details):

\begin{proposition}\label{Reconstruction2R1} The coefficients $\left(F_{\boldsymbol{\epsilon}^\star}^{\{k\}}\right)_{k\geq -2}$ are related to the symplectic invariants $\left(F^{(g)}\right)_{g\geq 0}$ computed from the topological recursion applied to the spectral curve \eqref{SpecCurve2R1} by:
\beaa \forall \,k\geq -1\,&:&\,\, F_{\boldsymbol{\epsilon}^\star}^{\{2k\}}=-F^{(2k+2)}+f_{2k} \text{ with } f_{2k} \text{ independent of } \epsilon \cr
 \forall \,k\geq -1\,&:&\,\, F_{\boldsymbol{\epsilon}^\star}^{\{2k+1\}}=f_{2k+1} \text{ with } f_{2k+1} \text{ independent of } \epsilon
 \eeaa
\end{proposition}

Note that the non-perturbative part of \eqref{GeneralForm2R1} starts at $O(1)$. The main difficulty of the expansion \eqref{GeneralForm2R1} lies in the fact that the non-perturbative part requires the knowledge of the spectral curve in the case where the filling fractions are arbitrarily fixed in order to compute the derivatives relatively to the filling fractions. In our case, if we take arbitrary filling fractions, the symmetries \eqref{Obs} are lost and thus the determination of the spectral curve requires to solve fixed filling fractions conditions:
\beqq \boldsymbol{\epsilon}_k=\oint_{\mathcal{A}_k}  d\mu_\infty(x) \text{  for all } -r\leq k\leq r \eeqq
with $\mathcal{A}_k$ a closed contour circling the interval $\left[\tan\left(\frac{\pi k}{2r+1}-\frac{\pi \epsilon}{2(2r+1)}\right),\tan\left(\frac{\pi k}{2r+1}+\frac{\pi \epsilon}{2(2r+1)}\right)\right]$. Unfortunately, solving analytically these conditions remains an open challenge and therefore the non-perturbative part of \eqref{GeneralForm2R1} remains mostly out of reach for theoretical computations. Another possibility to bypass this difficulty may be to rewrite derivatives relatively to filling fractions of $F^{\{2k\},(l)}$ as some integrals over $\mathcal{B}$-cycles of the correlation functions $\omega^{(k+1)}_l$ using results of \cite{EO}. In this case, there appears to have no need for using arbitrary filling fractions but the strategy requires the computations of all correlation functions. However this approach seems more convenient than dealing with arbitrary filling fractions. In particular, the case $m=0$ in \eqref{GeneralForm2R1} simply gives:
\beq \label{Termm0} \text{Term }m=0 \text{ in }\eqref{GeneralForm2R1}\,:\, \Theta_{-n\boldsymbol{\epsilon}^\star}\left(\mathbf{0}\big|F^{\{-2\},(2)}_{\boldsymbol{\epsilon}^\star}\right)=\Theta_{-n\boldsymbol{\epsilon}^\star}\left(\mathbf{0}\big|\,-2i\pi \boldsymbol{\tau}_{|\boldsymbol{\epsilon}=\boldsymbol{\epsilon}^*}\right)\eeq
where $\boldsymbol{\tau}$ is the Riemann matrix of periods of the spectral curve. In the case $\boldsymbol{\epsilon}=\boldsymbol{\epsilon}^*=\left(\frac{1}{2r+1},\dots,\frac{1}{2r+1}\right)^t$ the rotational symmetry of the problem may allow the computation of the Riemann matrix of periods even if the expression of the spectral curve is rather complicated (See \eqref{SpecCurve2R1}). Since we will have no use of this order, we let it as an open problem. 

\subsection{Normalization and limit $\epsilon \to 0$}

Finally, we also note that even the perturbative part is challenging in the multi-cut regime. Indeed, proposition \ref{Reconstruction2R1} determines the perturbative part up to some constants. The standard way to compute them is to choose the parameters (in our case $\epsilon$) in such a way that the initial integral is explicitly connected to a known case. In the one cut regime, we connected the integral to a Selberg integral after a proper rescaling. In the multi-cut case, we may also compute the limit of $Z_n(\mathcal{I}_r)$ when $\epsilon \to 0$ since the integrals decouple in this regime. We obtain the following theorem:

\begin{theorem}[Limit of $Z_n(\mathcal{I}_r)$ when $\epsilon\to 0$]\label{Theo2R1EpsilonTo0} For a given $r\geq 0$ and $n\geq 1$ we have:
\small{\bea \label{Zn2R1Limit}  Z_{n}(\mathcal{I}_r)&\overset{\epsilon \to 0}{\sim}&(2\pi\epsilon)^{(2r+1)\lfloor \frac{n}{2r+1}\rfloor^2+\left(2\lfloor \frac{n}{2r+1}\rfloor +1\right)m_n}\frac{(2\pi)^{-n}\left(\underset{k=1}{\overset{\lfloor \frac{n}{2r+1}\rfloor-1}{\prod}}k!\right)^{4(2r+1)} \left(\left(\lfloor \frac{n}{2r+1}\rfloor\right) !\right)^{4m_n}}{\left(\underset{k=1}{\overset{2\lfloor \frac{n}{2r+1}\rfloor-1}{\prod}}k!\right)^{2r+1}\left(\left(2\lfloor \frac{n}{2r+1}\rfloor\right)!\right)^{m_n} \left(\left(2\lfloor \frac{n}{2r+1}\rfloor+1\right)!\right)^{m_n}}\cr
&\overset{\epsilon \to 0}{\sim}&(2\pi\epsilon)^{\frac{n^2-m_n^2}{2r+1}+m_n}\frac{(2\pi)^{-n}\left(\underset{k=1}{\overset{\frac{n-m_n}{2r+1}-1}{\prod}}k!\right)^{4(2r+1)} \left(\left(\frac{n-m_n}{2r+1}\right) !\right)^{4m_n}}{\left(\underset{k=1}{\overset{\frac{2n-2m_n}{2r+1}-1}{\prod}}k!\right)^{2r+1}\left(\left(\frac{2n-2m_n}{2r+1}\right)!\right)^{m_n} \left(\left(\frac{2n-2m_n}{2r+1}+1\right)!\right)^{m_n}}\cr
&&\eea}
\end{theorem}

\normalsize{This} formula may easily be verified numerically by computing the leading order in the $\epsilon\to 0$ series expansion of the Toeplitz determinants given by \eqref{Toep2R1} for low values of $n$ (up to 25) and low values of $r$ (up to 10). Moreover, the formula reduces for $r=0$ to \eqref{CasRef}. When $r>1$, we observe as expected that the result depends both on the value of $n$ but also on the value of $n \mod (2r+1)$. This is of course in agreement with the asymptotic expansion given by \eqref{GeneralForm2R1}.

\medskip

\proof{
Let us first decompose the partition function as a sum over all possible partitions of the eigenvalues in the $2r+1$ intervals:
\footnotesize{\beaa Z_n(\mathcal{I}_r)&=&\frac{1}{(2\pi)^n n!}\sum_{\mathbf{n}\in N_{n,r}}\binom{n}{n_{-r},\dots, n_r}\int_{[\alpha_{-r}^{(r)},\beta_{-r}^{(r)}]} d\theta_1\dots d\theta_{n_{-r}}\dots \int_{[\alpha_{r}^{(r)},\beta_{r}^{(r)}]}d{\theta_{n-n_r+1}}\dots d{\theta_n}  \prod_{1\leq p<q\leq n}\left|e^{i\theta_p}-e^{i\theta_q}\right|^2\cr
&=&\frac{1}{(2\pi)^n n!}\sum_{\mathbf{n}\in N_{n,r}}\binom{n}{n_{-r},\dots, n_r}(2\pi)^{\underset{k=-r}{\overset{r}{\sum}} n_k}(n_{-r})!\,\dots (n_r)!\, Z_{n_{-r},\dots,n_r}(\mathcal{I}_r)\cr
&=&\sum_{\mathbf{n}\in N_{n,r}}Z_{n_{-r},\dots,n_r}(\mathcal{I}_r)
\eeaa}
\normalsize{where} $N_{n,r}=\{(i_{-r},\dots,i_r)\in \mathbb{N}^{2r+1}\,/\, \underset{k=-r}{\overset{r}{\sum}}i_k=n\}$, $\binom{n}{n_{-r},\dots, n_r}$ stands for the multinomial coefficient and 
\footnotesize{\beqq Z_{n_{-r},\dots,n_r}(\mathcal{I}_r)=\frac{1}{(2\pi)^{\underset{k=-r}{\overset{r}{\sum}} n_k}\underset{k=-r}{\overset{r}{\prod}} (n_k)!} \,\int_{[\alpha_{-r}^{(r)},\beta_{-r}^{(r)}]} d\theta_1\dots d\theta_{n_{-r}}\dots \int_{[\alpha_{r}^{(r)},\beta_{r}^{(r)}]}d{\theta_{n-n_r+1}}\dots d{\theta_n}  \prod_{1\leq p<q\leq n}\left|e^{i\theta_p}-e^{i\theta_q}\right|^2\eeqq}
\normalsize{stands} for the partition function with filling fractions fixed to $\mathbf{n}=(n_{-r},\dots,n_r)^t$.
Note that the normalization prefactor $\frac{1}{(2\pi)^n n!}$ in the definition of the partition functions exactly cancels the combinatorial coefficients appearing in the decomposition.

In the limit $\epsilon\to 0$, interactions between eigenvalues in two different intervals reduce to interactions between the center of the two intervals. Thus, the integrals decouple and $Z_{n_{-r},\dots,n_r}(\mathcal{I}_r)$ reduces to a product of one-cut partition functions whose values are given by \eqref{CasRef}  (with $a_0=\tan \frac{\pi \epsilon}{2(2r+1)}$):
\beq \label{Decoupling} Z_{n_{-r},\dots,n_r}(\mathcal{I}_r)\overset{\epsilon \to 0}{\sim} \underset{k=-r}{\overset{r}{\prod}} Z_{n_k}(\epsilon)=(2\pi\epsilon)^{\underset{k=-r}{\overset{r}{\sum}} n_k^2}(2\pi)^{-n} \frac{\underset{k=-r}{\overset{r}{\prod}}\left(\underset{j=1}{\overset{n_k}{\prod}} j!\right)^4}{\underset{k=-r}{\overset{r}{\prod}} \underset{j=1}{\overset{2n_k-1}{\prod}} j!}
\eeq

We now observe that the exponent in $\epsilon$ is minimal when the sum $\underset{k=-r}{\overset{r}{\sum}} n_k^2$ is minimal. Since the sum is constrained to $\underset{k=-r}{\overset{r}{\sum}}n_k=n$ fixed, it is well-known that the previous sum is minimal when all $(n_k)_{k\in\llbracket -r,r\rrbracket}$ are equal. This is also coherent with the fact that the perturbative analysis leads to identical filling fractions (See Proposition \ref{FillingFractions}). Indeed, let us define $f(\mathbf{n})=\underset{k=-r}{\overset{r}{\sum}} n_k^2$, then 
\beaa &&f(n_{-r},\dots,n_{i_0}+1,\dots,n_{j_0}-1,\dots,n_r)-f(n_{-r},\dots,n_{i_0},\dots,n_{j_0},\dots,n_r)\cr
&&=2(n_{i_0}-n_{j_0}+1)\left\{
    \begin{array}{ll}
        <0 \text{ if } n_{i_0}\leq n_{j_0}-2\\
        =0\text{ if } n_{i_0}=n_{j_0}-1\\
				>0 \text{ if } n_{i_0}\geq n_{j_0}
    \end{array}
\right.\eeaa
Therefore, to minimize $f$ on $N_{n,r}$, one needs to balance the vector $\mathbf{n}$ with differences between the components at most $1$. This proves that if $m_n=n\mod (2r+1)\neq 0$ the sum is minimal when we spread the additional $m_n$ eigenvalues in $m_n$ distinct intervals, i.e. we put $\lfloor \frac{n}{2r+1}\rfloor$ eigenvalues in $2r+1-m_n$ intervals and $\lfloor \frac{n}{2r+1}\rfloor +1$ eigenvalues in the remaining $m_n$ intervals and that all other configurations provide a strictly greater value of the exponent. Since the sum contains a finite number of terms ($n$ and $r$ are fixed so $N_{n,r}$ is a finite set) and because each of them is strictly positive (thus the sum of terms with the same exponent in $\epsilon$ cannot give of a strictly lower order in $\epsilon$), we obtain from \eqref{Decoupling} the result presented in Theorem \ref{Theo2R1EpsilonTo0}.
 }

\subsection{Final result and conjecture for $F^{\{0\}}$}
Note that the knowledge of  the equivalent of $Z_n(\mathcal{I}_r)$ when $\epsilon\to 0$ given by Theorem \ref{Theo2R1EpsilonTo0} is not sufficient to determine the constants $\left(f^{\{k\}}\right)_{k\geq -2}$ of Proposition \ref{Reconstruction2R1}. Indeed,  in order to obtain them, one needs to exchange the limit $\epsilon\to 0$ and the limit $n\to \infty$ and possible entropy contributions may arise. It is also unclear if arguments similar to the one-cut case may be adapted to this context. The only known cases are $F^{\{-2\}}$ and $F^{\{-1\}}$ 
(given in section $1.4$ of \cite{BG2} with the important reminder that we included a normalization factor $\frac{2^{n^2}}{(2\pi)^n n!}$ in the definition of the partition function that is absent in \cite{BG2}) . We get:

\beaa F^{\{-2\}}&=& \ln 2-F^{(0)} \,\,\,\Leftrightarrow \,\,\, f_{-2}=\ln 2  \cr
F^{\{-1\}}&=&0 \,\,\,\Leftrightarrow \,\,\, f_{-1}=0
\eeaa
\begin{remark}Note that the constants $f_{-2}=\ln 2$ and $f_{-1}=0$ differ from those given in \cite{BG2} because our choice of normalization of the partition function is different from the one used in \cite{BG2}. Indeed, our partition function is normalized with an additional factor $\frac{2^{n^2}}{(2\pi)^n n!}$ compared to the one used in \cite{BG2}. In particular, the numerator $2^{n^2}$ provides an additional factor $n^2\ln 2$ in the asymptotic expansion of $\ln Z_n(\mathcal{I}_r)$ thus giving $f_{-2}=\ln 2$. Similarly, the denominator $(2\pi)^nn!\overset{n\to \infty}{\sim}(2\pi)^nn^ne^{-n}\sqrt{2\pi n}$ provides an additional factor $n\ln n$ already included in the general form of the asymptotic expansion of $\ln Z_n(\mathcal{I}_r)$ given in Proposition \ref{PropBorot} as well as an additional factor $-n(\ln (2\pi)-1)$ in the asymptotic expansion of $\ln Z_n(\mathcal{I}_r)$ that precisely cancels the one obtained in \cite{BG2}, hence leading to $f_{-1}=0$.
\end{remark}

Thus, we end up with:
\begin{proposition}[First orders of the asymptotic expansion of $Z_n(\mathcal{I}_r)$] For a given $r\geq 0$, we have: 
\beqq \ln (Z_n(\mathcal{I}_r))=\frac{n^2}{2r+1}\ln \left(\sin\left(\frac{\pi \epsilon}{2}\right)\right)-\frac{2r+1}{4}\ln n+O(1)
\eeqq 
\end{proposition}

As explained above, the next orders of the large $n$ expansion, and in particular the $O(1)$ term, exhibit a far more complex structure than in the one-cut case and depend on the remainder of the Euclidean division of $n$ by $2r+1$. In fact, as discussed in \cite{BG2}, for a given remainder $m$ such that $0\leq m \leq 2r$, the sequence $\left(Z_{(2r+1)p+m}\right)_{p\in \mathbb{N}}$ admit a large $p$ expansion taking a similar form as the one-cut case with $n$ replaced by $p$ (the precise form being given in Theorem \ref{LargeNExp}) and with coefficients depending on the remainder $m$. However, because we could compute numerically the Toeplitz determinants up to sufficiently large $n$, we can propose some conjectures for the coefficient $F^{\{0\}}$ depending on the value of the remainder $m$. 

\begin{conjecture}[Conjecture for $F^{\{0\}}$]\label{Conj} We conjecture the following:
\begin{itemize}\item In the case $m=0$, $F^{\{0\}}$ has a dependence in $\epsilon$ given by $-\frac{2r+1}{4}\ln\left(\cos\left(\frac{\pi \epsilon}{2}\right)\right)+C_{0}$ where $C_{0}$ is a constant. This generalizes the one-cut case that would correspond to $r=0$.
\item The cases $m=j$ and $m=2r-1-j$ for some $1\leq j\leq r$ provides the same $F^{\{0\}}$ coefficient. However, the dependence in $\epsilon$ is more involved than for $m=0$ and we conjecture that it is given by:
\beqq F^{\{0\}}(j)= A_{j}\ln\left(\cos\left(\frac{\pi \epsilon}{2}\right)\right)+B_{j} \ln\left(\tan\left(\frac{\pi \epsilon}{2}\right)\right) +C_j\eeqq 
with some non-zero constants $(A_j,B_j,C_j)$ depending on the remainder.
\item The constants $(C_j)_{1\leq j\leq r}$ and $C_{0}$ are the same and correspond to the normalization issue of the partition function.
\end{itemize}
\end{conjecture}

The case $m=0$ exhibits an additional symmetry since the number of eigenvalues is precisely a multiple of the number of intervals so that they can spread evenly in all intervals. In particular when evaluating derivatives of the free energies at $\boldsymbol{\epsilon}^\star=\frac{1}{2r+1}\mathbf{1}$ cancellations are more likely to happen, thus explaining why the coefficient in front of $\ln\left(\tan\left(\frac{\pi \epsilon}{2}\right)\right)$ vanishes only in this case. 
We illustrate our conjecture with numeric simulations:

\begin{center}
\includegraphics[width=16cm]{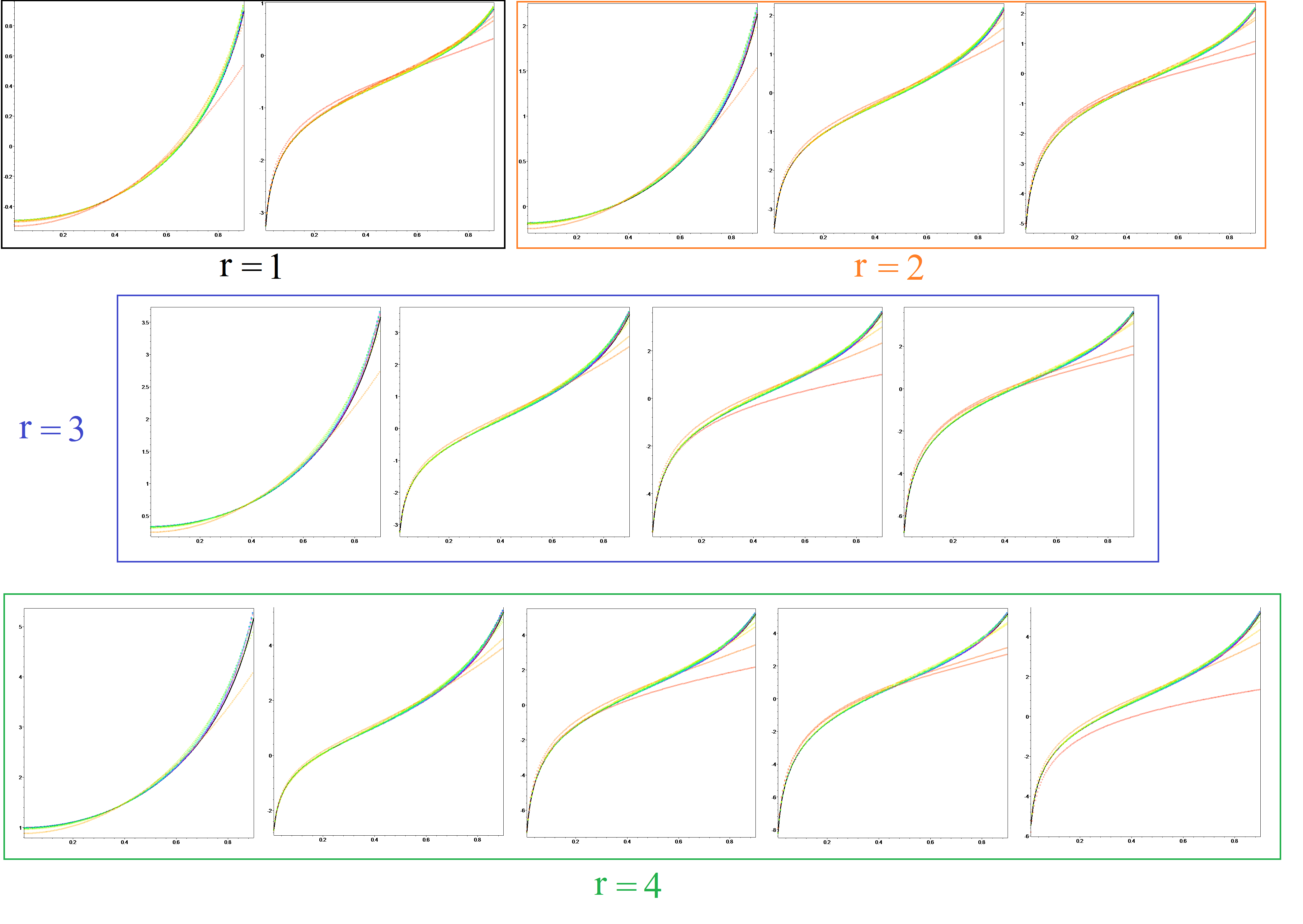}

Fig. $6$: Representation of $\ln (Z_n(\mathcal{I}_r)) -\frac{1}{2r+1}\ln \left(\sin\left(\frac{\pi \epsilon}{2}\right)\right)+\frac{2r+1}{4}\ln n$ for values of $n$ ranging from $2$ to $70$ and classified by the remainder $m$ (from left to right: $m=0$, $m\in\{j,2r-1-j\}$, with $j$ from $1$ to $r$) of the Euclidean division of $n$ by $2r+1$. The black curves correspond to the best numerical matches of an affine combination of $\ln\left(\cos\left(\frac{\pi \epsilon}{2}\right)\right)$ and $\ln\left(\tan\left(\frac{\pi \epsilon}{2}\right)\right)$ as explained more specifically below.
\end{center}

More precisely, the black curves correspond to the best matches with curves of the form:
\beq f(\epsilon)=\alpha\ln\left(\cos\left(\frac{\pi \epsilon}{2}\right)\right) +\beta \ln\left(\tan\left(\frac{\pi \epsilon}{2}\right)\right)+\gamma\eeq
with rational coefficients $(\alpha,\beta,\gamma)$ of the form $\frac{i}{256}$ with $i\in \mathbb{Z}$. Note that we chose to express the coefficients as rational numbers with a specific denominator, but we have no evidence that the coefficients are indeed rational or that the denominator is a power of $2$. However it seems that this particular choice provides very accurate results that are not numerically improved by adding multiplicative powers of $3$ or $5$ in the denominators or increasing the power of $2$. Numerically, we obtain the best matches for the values:
\begin{itemize}\item \underline{Case $r=1$}:
\beaa m=0 &:& (\alpha,\beta,\gamma)= \left(-\frac{3}{4},0,-\frac{63}{128}\right)\cr
m\in\{1,2\}&:& (\alpha,\beta,\gamma)= \left(-\frac{11}{128},\frac{85}{128},-\frac{63}{128}\right)
\eeaa
\item \underline{Case $r=2$}:
\beaa m=0 &:& (\alpha,\beta,\gamma)= \left(-\frac{5}{4},0,-\frac{3}{16}\right)\cr 
m\in\{1,4\}&:& (\alpha,\beta,\gamma)= \left(-\frac{15}{32},\frac{203}{256},-\frac{3}{16}\right)\cr
m\in\{2,3\}&:& (\alpha,\beta,\gamma)= \left(-\frac{1}{16},\frac{6}{5},-\frac{3}{16}\right)
\eeaa
\item \underline{Case $r=3$}:
\beaa m=0 &:& (\alpha,\beta,\gamma)=\left(-\frac{7}{4},0,\frac{85}{256}\right)\cr
m\in\{1,6\}&:&(\alpha,\beta,\gamma)=\left(-\frac{115}{128},\frac{109}{128},\frac{85}{256}\right)\cr
m\in\{2,5\}&:&(\alpha,\beta,\gamma)=\left(-\frac{43}{128},\frac{183}{128},\frac{85}{256}\right)\cr 
m\in\{3,4\}&:&(\alpha,\beta,\gamma)=\left(-\frac{7}{128},\frac{219}{128},\frac{85}{256}\right) 
\eeaa 
\item \underline{Case $r=4$}:
\beaa m=0 &:& (\alpha,\beta,\gamma)=\left(-\frac{9}{4},0,\frac{255}{256}\right)\cr
m\in\{1,8\}&:&(\alpha,\beta,\gamma)=\left(-\frac{355}{256},\frac{227}{256},\frac{255}{256}\right)\cr
m\in\{2,7\}&:&(\alpha,\beta,\gamma)=\left(-\frac{184}{256},\frac{398}{256},\frac{255}{256}\right)\cr 
m\in\{3,6\}&:&(\alpha,\beta,\gamma)=\left(-\frac{72}{256},\frac{510}{256},\frac{255}{256}\right)\cr 
m\in\{4,5\}&:&(\alpha,\beta,\gamma)=\left(-\frac{17}{256},\frac{565}{256},\frac{255}{256}\right) 
\eeaa 
\end{itemize}


\subsection{Even number of intervals \label{SectionEvenCase}}

The method developed in the last section can be adapted in the case of an even number of intervals. However we need to be careful since in order to apply $\theta \mapsto \tan \frac{\theta}{2}$ we need to avoid the angles $\theta=\pm \pi$. Therefore we use the invariance $\boldsymbol{\theta}\to \boldsymbol{\theta}+\text{Cste }\,\mathbf{1}$ of the integral \eqref{EvenCase} to shift the intervals so that they do not contain $\pm \pi$. We define for $s\geq 1$:
\beaa \alpha_k^{(s)}&=&\frac{\pi (k-\frac{1}{2})}{s}- \frac{\pi \epsilon}{2s} \text{ ,  } \beta_k^{(s)}=\frac{\pi (k-\frac{1}{2})}{s}+ \frac{\pi \epsilon}{2s} \text{ , } \gamma_k^{(s)}=\frac{\pi (k-\frac{1}{2})}{s} \,\,\text{ , } \,\, \forall\, -(s-1)\leq k\leq s\cr 
\mathcal{I}_s&=&\underset{k=-(s-1)}{\overset{s}{\bigcup}}\left[\alpha_k^{(s)},\beta_k^{(s)}\right]\cr
\mathcal{J}_s&=&\underset{k=-(s-1)}{\overset{s}{\bigcup}}\left[\tan\left(\frac{\alpha_k^{(s)}}{2}\right),\tan\left(\frac{\beta_k^{(s)}}{2}\right)\right]\cr
\mathcal{T}_s&=&\{e^{it}\,,\,\, t\in \mathcal{I}_s\}
\eeaa
and the integral:
\beq \label{EvenCase} Z_n(\mathcal{I}_s)=\frac{1}{(2\pi)^n n!} \int_{\left(\mathcal{I}_s\right)^n} d\theta_1\dots d\theta_n \prod_{1\leq p<q\leq n}\left|e^{i\theta_p}-e^{i\theta_q}\right|^2 \eeq
which is also equal to $\det T_n^{(s)}$ with $\left(T_n^{(s)}\right)_{p,q}=t_{p-q}$ the $n\times n$ Toeplitz matrix given by:
\beaa t_0&=&\frac{|\mathcal{I}_s|}{2\pi}=\epsilon\cr
t_k&=& \epsilon \, \sin_c\left(\frac{k \pi \epsilon}{2s}\right) \delta_{k \, \equiv \, 0\,[2s]} \,\text{ for }k\neq 0
\eeaa
Note again that the Toeplitz matrix is mostly empty since only bands with indexes multiple of $2s$ are non-zero. Moreover it is also equal to a Hermitian integral: 
\beq \label{RealEven} Z_n(\mathcal{I}_s)=\frac{2^{n^2}}{(2\pi)^n n!}\int_{\left(\mathcal{J}_s\right)^n} dt_1\dots dt_n  \Delta(t_1,\dots,t_n)^2e^{-n\underset{k=1}{\overset{n}{\sum}} \ln(1+t_k^2)}\eeq
or a complex integral:
\beq \label{ComplexREven} Z_n(\mathcal{I}_s)=(-1)^{\frac{n(n+1)}{2}}i^n \int_{\left(\mathcal{T}_s\right)^n} du_1\dots d u_n\, \Delta(u_1,\dots,u_n)^2 e^{-n\underset{k=1}{\overset{n}{\sum}} \ln u_k}\eeq 

The situation can be illustrated as follow:
\begin{center}
\includegraphics[width=6cm]{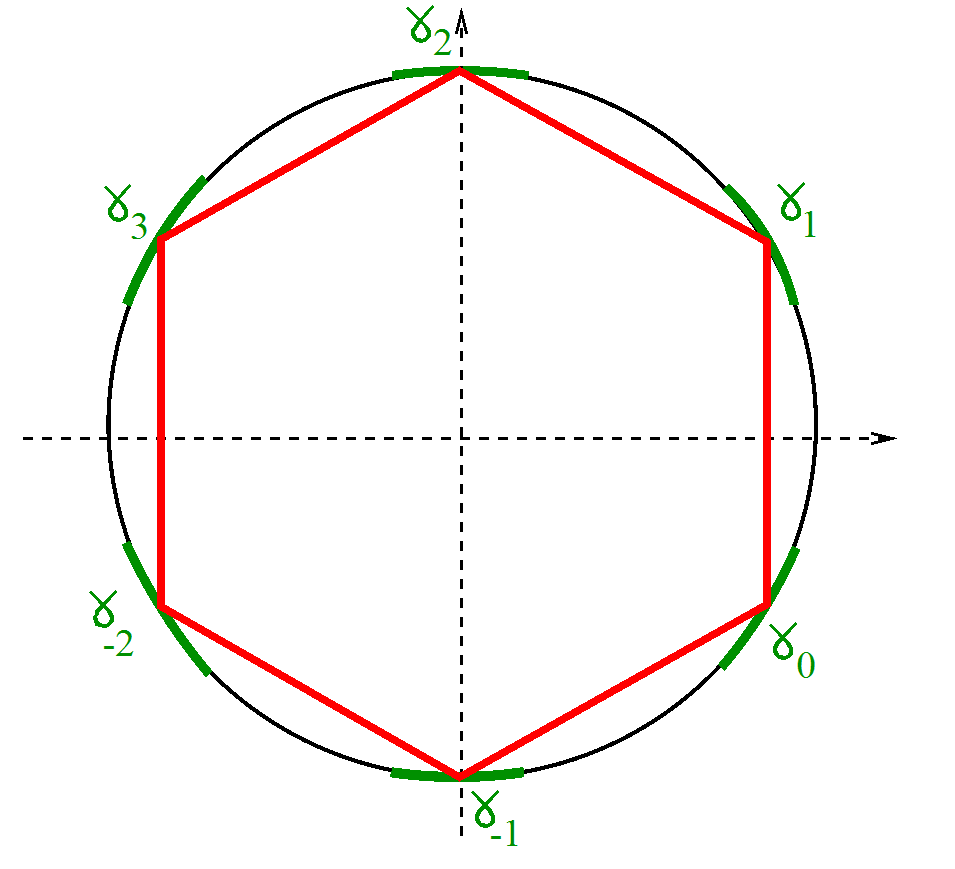}

Fig. $7$: Illustration (in green) of the set $\mathcal{T}_{s=3}$ for $\epsilon=\frac{1}{5}$.
\end{center}

The method developed in the case of an odd number of intervals can be easily adapted to this case. In particular we can prove the following:

\begin{theorem}[Results for an even number of symmetric arc-intervals] In the case of an even number of symmetric arc-intervals \eqref{EvenCase} we have:
\begin{enumerate}\item The spectral curve attached to integral \eqref{EvenCase} is given by:
\beaa 
y^2(x)
&=&\frac{\underset{k=1}{\overset{s-1}{\prod}} \cos^2\left(\frac{\pi\left(k-\frac{1}{2}\right)}{2s}\right)}{\underset{k=-(s-1)}{\overset{s}{\prod}} \cos^2\left(
\frac{\pi\left(k-\frac{1}{2}\right)}{2s}-\frac{\pi\epsilon}{4s}\right)} \frac{ x^2\underset{k=1}{\overset{s-1}{\prod}}\left(x^2-\tan^2\left(\frac{\pi k}{2s}\right)\right)^2}{(1+x^2)^2\underset{k=-(s-1)}{\overset{s}{\prod}}\left(x^2-\tan^2\left(\frac{\pi\left(k-\frac{1}{2}\right)}{2s}+\frac{\pi \epsilon}{4s}\right)\right)}\cr
\eeaa
\item The corresponding limiting eigenvalues distribution is given by:
\small{\beaa d\mu_\infty(x)=\frac{\underset{k=1}{\overset{s-1}{\prod}} \cos\left(\frac{\pi\left(k-\frac{1}{2}\right)}{2s}\right)}{\underset{k=-(s-1)}{\overset{s}{\prod}} \cos\left(
\frac{\pi\left(k-\frac{1}{2}\right)}{2s}-\frac{\pi\epsilon}{4s}\right)} \frac{ |x|\underset{k=1}{\overset{s-1}{\prod}}\left|x^2-\tan^2\left(\frac{\pi k}{2s}\right)\right|}{\pi(1+x^2)\underset{k=-(s-1)}{\overset{s}{\prod}}\sqrt{\left|x^2-\tan^2\left(\frac{\pi\left(k-\frac{1}{2}\right)}{2s}+\frac{\pi \epsilon}{4s}\right)\right|}}\mathds{1}_{\mathcal{J}_s}(x)dx\cr 
\eeaa}\normalsize{}
\item The filling fractions are the same in all the $2s$ intervals:
\beqq \forall\, k\in\llbracket -(s-1),s\rrbracket \,:\, \boldsymbol{\epsilon}_{k+s}\overset{\text{def}}{=}\int_{\tan\left(\frac{\pi\left(k-\frac{1}{2}\right)}{2s}-\frac{\pi \epsilon}{4s}\right)}^{\tan\left(\frac{\pi\left(k-\frac{1}{2}\right)}{2s}+\frac{\pi \epsilon}{4s}\right)} d\mu_\infty(x)=\frac{1}{2s}\eeqq
\item The function $\ln Z_n(\mathcal{I}_s)$ admits a large $n$ expansion given by Theorem \ref{PropBorot} with partial reconstruction by the topological recursion given by proposition \ref{Reconstruction2R1}. In particular we have for $s\geq 1$:
\beqq\ln (Z_n(\mathcal{I}_s))=\frac{1}{2s}\ln \left(\sin\left(\frac{\pi \epsilon}{2}\right)\right)-\frac{2s}{4}\ln n+O(1)\eeqq
The $O(1)$ order depends on the remainder $m_n$ of $n$ modulo $2s$. Numerically, its dependence in $\epsilon$ follows the conjecture \ref{Conj} (with $2r+1$ replaced by $2s$ and the special case still given by $m_n=0$). 
\item The limit $\epsilon \to 0$  of $Z_n(\mathcal{I}_s)$ with $n=m_n\mod 2s$ is given by:
\small{\beq \label{Zn2RLimit}  Z_{n}(\mathcal{I}_s)\overset{\epsilon \to 0}{\sim}(2\pi\epsilon)^{2s\lfloor \frac{n}{2s}\rfloor^2+\left(2\lfloor \frac{n}{2s}\rfloor+1\right)m_n}\frac{(2\pi)^{-n}\left(\underset{k=1}{\overset{\lfloor \frac{n}{2s}\rfloor-1}{\prod}}k!\right)^{8s} \left(\left(\lfloor \frac{n}{2s}\rfloor\right) !\right)^{4m_n}}{\left(\underset{k=1}{\overset{2\lfloor \frac{n}{2s}\rfloor-1}{\prod}}k!\right)^{2s}\left(\left(2\lfloor \frac{n}{2s}\rfloor\right)!\right)^{m_n} \left(\left(2\lfloor \frac{n}{2s}\rfloor+1\right)!\right)^{m_n}}
\eeq}
\normalsize{}
\end{enumerate} 
\end{theorem}

\section{Conclusion and outlooks}

In this article, we proposed a rigorous mathematical derivation of the asymptotic expansions of Toeplitz determinants with symbols given by $f=\mathds{1}_{\mathcal{T}(\alpha,\beta)}$ where $\mathcal{T}(\alpha,\beta)=\{e^{it},t\in [\alpha,\beta]\}$. This generalizes Widom's result and completes the approach developed in \cite{UnitaryMarchal}. We also provided numerical simulations up to $o\left(\frac{1}{n^4}\right)$ to illustrate these results. For symbols $f=\mathds{1}_{\mathcal{T}_d}$ with $\mathcal{T}_d=\underset{k=1}{\overset{d}{\bigcup}}[\alpha_k,\beta_k]$ and $d\geq 2$, the situation is more complex, but we were able to provide a rigorous derivation of the large $n$ asymptotic of the corresponding Toeplitz determinants when the arc-intervals exhibit a discrete rotational symmetry on the unit circle. We also provided the first terms of the large $n$ expansion up to $O(1)$ and we proposed a conjecture for the $O(1)$ term supported by numeric simulations. Moreover, the results presented in this article raise the following challenges:
\begin{itemize} \item Prove conjecture \ref{Conj} regarding the $O(1)$ term in the symmetric multi-cut case. In particular, it would be interesting to find a way to obtain the normalization constants. This requires to connect the Toeplitz integral to a known integral for at least one value of the parameter $\epsilon$ (very likely $\epsilon\to 0$).
\item In the multi-cut cases without discrete symmetry, it would be interesting to prove that, for generic choices of the edges, the corresponding spectral curves are regular and that the hypothesis required to prove the large $n$ asymptotic of Theorem \ref{PropBorot} are verified. Contrary to the case with a discrete rotational symmetry, it seems unlikely that we obtain an explicit formula for the spectral curves. However it may be possible to obtain sufficient information on the location of the zeros of $y(x)$ in order to prove that the spectral curves are regular.
\item The method developed in this article could also be tried for more general Toeplitz determinants. In particular, the transition from a symbol supported on a compact set of $\{e^{it},t\in(-\pi,\pi)\}$ to a strictly positive symbol on the unit circle deserves some analysis. Indeed, in the case of a strictly positive symbol on the unit circle, Szeg\"{o}'s theorem implies that $\frac{1}{n}\ln \det T_n$ admits a non-trivial limit (given by $\frac{1}{2\pi}\int_0^{2\pi} \ln(f(e^{i\theta}))d\theta$) so that the large $n$ expansion proposed in Theorem \ref{PropBorot} requires some adaptations. However, it may happen that the sub-leading corrections may still be given by Theorem \ref{PropBorot} and in particular that sub-leading corrections may be reconstructed from the topological recursion. This conjecture is supported by the fact that Toeplitz determinants can also be reformulated as Fredholm determinants (See \cite{Fredholm}) that are known to be deeply related with the topological recursion.     
\end{itemize}   

\section*{Acknowledgments}
I would like to thank A. Guionnet that indirectly gave me the idea to review in a rigorous way the tools and theorems used in this article. I also would like to thank G. Borot for very fruitful discussions about the normalizing constants and also an unknown referee for his suggestions. Eventually, I would like to thank Universit\'e de Lyon, Universit\'e Jean Monnet and Institut Camille Jordan for financial and material support.

\begin{appendix}

\section{\label{TopRec1Cut}Topological recursion and normalization for the one-cut case}
In this appendix, we provide details about the computation of the free energies $\left(F_{\text{Top. Rec.}}^{(g)}(\epsilon)\right)_{g\geq 0}$ required for section \ref{Section1Int}. Computations are direct application of the topological recursion whose general definitions and results are presented in Appendix \ref{AppendixTopRec}.

\subsection{Computation of the topological recursion\label{DirectAppliTopRec}}
In this section we present the computation of the topological recursion to the spectral curve: 
\bea \label{Specc} x(z)&=&\frac{1}{2}\tan\left(\frac{\pi \epsilon}{2}\right)\left(z+\frac{1}{z}\right)\cr
y(z)&=&\frac{2}{\sin\left(\frac{\pi \epsilon}{2}\right)\left(1+\frac{1}{4}\tan^2\left(\frac{\pi \epsilon}{2}\right)\left(z+\frac{1}{z}\right)\right)\left(z-\frac{1}{z}\right)}
\eea
The spectral curve is of genus $0$ and the branchpoints are located at $z=\pm 1$ and we can define a global involution $\bar{z}=\frac{1}{z}$ for which $x(\bar{z})=x(z)$ and $y(\bar{z})=-y(z)$. Note that $y$ has simple poles at the branchpoints so that the one-form $ydx$ given by:
\beq \label{ydx} ydx(z)=\frac{dz}{z\cos(\frac{\pi \epsilon}{2})\left(1+\frac{1}{4}\tan^2(\frac{\pi \epsilon}{2})\left(z+\frac{1}{z}\right)\right)}\eeq
is regular at the branchpoints. We now need to compute the first free energies $\left(F_{\text{Top. Rec.}}^{(g)}(\epsilon)\right)_{g\geq 0}$ attached to this genus $0$ spectral curve using the formulas presented in Appendix \ref{AppendixTopRec}. Specific formulas presented in \cite{EO} are required for $F_{\text{Top. Rec.}}^{(0)}(\epsilon)$ and $F_{\text{Top. Rec.}}^{(1)}(\epsilon)$ and the corresponding computations are presented in Appendix \ref{AppendixF0F1} (equations \eqref{F0Int} and \eqref{F1Int}). We find (remind that $a=\tan \frac{\pi \epsilon}{2}$):

\bea \label{ResultTopReccc}F_{\text{Top. Rec.}}^{(0)}(\epsilon)&=&\ln 2-\ln\left(\sin\frac{\pi \epsilon}{2}\right)  \cr
F_{\text{Top. Rec.}}^{(1)}(\epsilon)&=&\frac{1}{4}\ln\left(\cos\left(\frac{\pi \epsilon}{2}\right)\right) \cr
F_{\text{Top. Rec.}}^{(2)}(\epsilon)&=&\frac{1}{64}-\frac{1}{32}\tan^2\left(\frac{\pi \epsilon}{2}\right)\cr
F_{\text{Top. Rec.}}^{(3)}(\epsilon)&=&-\frac{1}{256}-\frac{1}{128}\tan^2\left(\frac{\pi \epsilon}{2}\right)-\frac{5}{128}\tan^4\left(\frac{\pi \epsilon}{2}\right)
\eea

\begin{remark}Note that we have $W_{p,a}^{(0)}(x_1,\dots,x_p)=0$ for all $p\geq 3$ because $y$ has a simple pole at the branchpoints. Moreover, from the definition and the form of the spectral curve, it is also easy to see that the coefficients $\left(W_{p,a}^{(g)}(x_1,\dots,x_p)\right)_{p\geq 1, g\geq 0}$ and $\left(F_{\text{Top. Rec.}}^{(g)}(a)\right)_{g\geq 0}$ are polynomial functions of $a=\tan(\frac{\pi \epsilon}{2})$ and $\sqrt{1+a^2}=\frac{1}{\cos(\frac{\pi \epsilon}{2})}$.
\end{remark}

\subsection{Computation of $F^{(0)}_{\text{Top.Rec}}$ and $F^{(1)}_{\text{Top.Rec}}$ \label{AppendixF0F1}}
We want to compute the first two free energies of the curve:
\bea x(z)&=&\frac{1}{2}\tan\left(\frac{\pi \epsilon}{2}\right)\left(z+\frac{1}{z}\right)\cr
y(z)&=&\frac{2}{\sin(\frac{\pi \epsilon}{2})\left(1+\frac{1}{4}\tan^2(\frac{\pi \epsilon}{2})\left(z+\frac{1}{z}\right)\right)\left(z-\frac{1}{z}\right)}
\eea
We first observe that the one-form $ydx$ is given by:
\beq ydx(z)=\left(\frac{t_1}{z-Z_1}+\frac{t_2}{z-Z_2}+\frac{t_3}{z-Z_3}+\frac{t_4}{z-Z_4} \right)dz\eeq
with:
\bea Z_1&=&i\frac{1-\cos(\frac{\pi \epsilon}{2})}{\sin(\frac{\pi \epsilon}{2})}  \text{  and  } t_1=\frac{1}{2}\cr
Z_2&=&i\frac{1+\cos(\frac{\pi \epsilon}{2})}{\sin(\frac{\pi \epsilon}{2})} \text{  and  } t_2=-\frac{1}{2}\cr
Z_3&=&-i\frac{1+\cos(\frac{\pi \epsilon}{2})}{\sin(\frac{\pi \epsilon}{2})} \text{  and  } t_3=-\frac{1}{2}\cr
Z_4&=&-i\frac{1-\cos(\frac{\pi \epsilon}{2})}{\sin(\frac{\pi \epsilon}{2})}  \text{  and  } t_1=\frac{1}{2}
\eea
Thus the one form $ydx$ has $4$ simple poles that are not branchpoints but only poles of $y(z)$. The local coordinate around each point is given by:
\beqq x_k(z)=\frac{1}{x(z)-x(Z_k)}=\frac{2z}{\tan(\frac{\pi \epsilon}{2}) (z-Z_k)(z-\frac{1}{Z_k})}\eeqq 
Hence the local potential:
\beqq V_k(p)=\Res_{q\to Z_k} ydx(q)\ln\left(1-\frac{x(q)-x(Z_1)}{x(p)-x(Z_1)}\right)\eeqq
is trivially vanishing in all four cases since the poles are simple. Eventually we end up with the computation of:
\beq \mu_k=\left(\int^o_{Z_k}ydx(z)-t_k\frac{dx(z)}{x(z)-x(Z_k)}\right)+t_k\ln(x(r)-x(Z_k))\eeq
Observing that $\frac{x'(z)}{x(z)-x(Z_k)}=-\frac{1}{z}+\frac{1}{z-Z_k}+\frac{1}{z-\frac{1}{Z_k}}$ we have:
\beq \mu_k=\sum_{j\neq k}t_j\ln(o-Z_j)-\sum_{j\neq k} t_j\ln(Z_k-Z_j) +t_k\ln\left(1-\frac{1}{Z_k^2}\right)+\frac{t_k}{2}\tan\left(\frac{\pi \epsilon}{2}\right)
\eeq
Hence in the end, observing that $\underset{k=1}{\overset{4}{\sum}} t_j=0$ we get that the dependence in $o$ vanishes (as claimed in \cite{EO}) and we find:
\bea \label{F0Int} F^{(0)}_{\text{Top.Rec.}}&\overset{\text{def}}{=}&\frac{1}{2}\sum_{k=1}^4 t_k\mu_k\cr
&=&-\frac{1}{2}\sum_{k=1}^4\sum_{j\neq k}t_kt_j\ln(Z_k-Z_j)+\frac{1}{2}\sum_{k=1}^4 t_k\ln\left(1-\frac{1}{Z_k^2}\right)+\frac{1}{2}\tan\left(\frac{\pi \epsilon}{2}\right)\sum_{k=1}^4t_k^2\cr
&=&-\frac{1}{2}\sum_{j<k=1}^4t_kt_j\ln(-(Z_k-Z_j)^2)+\frac{1}{2}\sum_{k=1}^4 t_k\ln\left(1-\frac{1}{Z_k^2}\right)+\frac{1}{2}\tan\left(\frac{\pi \epsilon}{2}\right)\sum_{k=1}^4t_k^2\cr
&=&\ln 2-\ln\left(\sin \frac{\pi \epsilon}{2}\right)
\eea

\bigskip

The computation of $F^{(1)}_{\text{Top.Rec.}}$ in the case of two hard edges is relatively straightforward. First, we observe that we have:
\beq \label{W1111}W_1^{(1)}(z)=\frac{1}{2\cos\left(\frac{\pi \epsilon}{2}\right)(z-1)^2(z+1)^2}\eeq
Then the computation of $F^{(1)}_{\text{Top.Rec.}}$ can be performed with the formalism of \cite{HardWall} that includes the possibility of hard edges. Straightforward, but lengthy, computations using equation $[149]$ of \cite{HardWall} gives:
\beq \label{F1Int} F^{(1)}_{\text{Top.Rec.}}=\frac{1}{4}\ln\left(\cos\left(\frac{\pi \epsilon}{2}\right)\right)=-\frac{1}{8}\ln\left(1+\tan^2\left(\frac{\pi \epsilon}{2}\right)\right)\eeq
An alternative approach with detailed computations using a refinement of the formalism of \cite{EO} adapted to hard edges is developed in \cite{UnitaryMarchal}. However, we note that Bergman tau-function used in \cite{UnitaryMarchal} is slightly incorrect. Indeed since $W_1^{(1)}(z)=\frac{1}{2\cos\left(\frac{\pi \epsilon}{2}\right)(z-1)^2(z+1)^2}$, then the Bergman tau-function in the present case is $\ln \tau_B= \frac{3}{8}\ln\left(\cos\left(\frac{\pi \epsilon}{2}\right)\right)$.

\subsection{Normalization with a Selberg integral \label{Normalization1Cut}}
\begin{remark}In this section, we will use the term ``asymptotic expansion'' and infinite series in the same sense as Theorem \ref{LargeNExp}. In particular, infinite series like $f(N)=\underset{k=1}{\overset{\infty}{\sum}} a_k N^{-k}$ are to be understood as $f(N)=\underset{k=1}{\overset{K}{\sum}}  a_k N^{-k} +o\left(N^{-K}\right)$ for all $K\geq 1$.
\end{remark}

We have the following Selberg integral:
\bea \label{Selberg} S_n(1,1,1)&=&\int_{[-1,1]^n} \prod_{1\leq i<j\leq n} (u_i-u_j)^2du_1\dots du_n=\frac{2^{n^2}}{n!}\prod_{j=1}^{n-1}\frac{ \Gamma^2(j+1)\Gamma(j+2)}{\Gamma(n+j+1)\Gamma(2)}\cr
&=&\frac{2^{n^2}(n!)\left(\underset{j=1}{\overset{n-1}{\prod}}j!\right)^4}{\underset{j=1}{\overset{2n-1}{\prod}}j!}\eea
We get:
\bea \label{CasRef} Z_n(a_0)&=&\frac{2^{n^2}}{(2\pi)^n n!}\int_{[-a_0,a_0]^n} \Delta(t_1,\dots,t_n)^2 e^{-n\underset{k=1}{\overset{n}{\sum}}\ln(1+t_k^2)}dt_1\dots dt_n\cr
&=&\frac{a_0^{n^2}2^{n^2}}{(2\pi)^n n!}\int_{[-1,1]^n} \Delta(u_1,\dots,u_n)^2e^{-n\underset{k=1}{\overset{n}{\sum}}\ln(1+a_0^2u_k^2)}du_1\dots du_n\cr
&\overset{\text{def}}{=}&\frac{a_0^{n^2}2^{n^2}}{(2\pi)^n n!}S_{a_0}
\eea
with:
\beq S_{a_0}=\int_{[-1,1]^n} \Delta(u_1,\dots,u_n)^2e^{-n\underset{k=1}{\overset{n}{\sum}}\ln(1+a_0^2u_k^2)}du_1\dots du_n\eeq
$S_{a_0}$ is a Hermitian matrix integral on $I=[-1,1]$ with potential $V_{a_0}(x)=\ln(1+a_0^2x^2)$. It is continuous in $a_0$ and in particular for $a_0=0$ we find:
\beq S_0=S_n(1,1,1)=\frac{2^{n^2}(n!)\left(\underset{j=1}{\overset{n-1}{\prod}}j!\right)^4}{\underset{j=1}{\overset{2n-1}{\prod}}j!}\eeq
Therefore, we have:
\bea \label{NormalizationEq}\ln Z_n(a_0)-n^2\ln(a_0)&\underset{a_0\to 0}{\to}& \ln\left(S_n(1,1,1)\right)+2n^2\ln 2-n\ln(2\pi)-\ln(n!)\cr
&=&4\ln (G(n+1))-\ln (G(2n+1))+2n^2\ln 2-n\ln(2\pi)\cr
&&\eea
where the function $G(z)$ is the $G$-Barnes function whose asymptotic expansion is:
\bea &&\label{Barnes} \ln (G(N+1))=\ln \left(\prod_{j=1}^{N-1}j!\right)\cr
&&=\frac{N^2}{2}\ln N+\frac{N}{2}\ln(2\pi) -\frac{1}{12}\ln N +\zeta'(-1)-\frac{3}{4}N^2+\sum_{g=2}^{\infty} \frac{B_{2g}}{2g(2g-2)} N^{2-2g}
\eea
Using this formula for $N=n$ and $N=2n$ as well as Stirling's asymptotic formula:
\beq \ln(n!)=\frac{1}{2}\ln(2\pi)+n\ln n+\frac{1}{2}\ln n -n+\sum_{k=1}^\infty \frac{B_{2k}}{2k(2k-1)n^{2k-1}}\eeq
we can compute the asymptotic expansion of the r.h.s. of \eqref{NormalizationEq}. We obtain:
\bea \label{Side1}&&\ln\left(S_n(1,1,1)\right)+2n^2\ln 2-n\ln(2\pi)-\ln(n!)\cr
&&=-\frac{1}{4}\ln n +3\zeta'(-1)+\frac{1}{12}\ln 2 +\sum_{g=1}^\infty \frac{4(1-2^{-2g-2})B_{2g+2}}{2g(2g+2) n^{2g}}\eea
so that:
\bea \label{SelbergLimit} \ln Z_n(a_0)-n^2\ln(a_0)&\underset{a_0\to 0}{\to}&4\ln (G(n+1))-\ln (G(2n+1))+2n^2\ln 2-n\ln(2\pi)\cr
&=&-\frac{1}{4}\ln n +3\zeta'(-1)+\frac{1}{12}\ln 2 +\sum_{g=1}^\infty \frac{4(1-2^{-2g-2})B_{2g+2}}{2g(2g+2) n^{2g}}\cr
&&
\eea

\subsection{Term by term limit of the free energies when $\epsilon\to 0$ \label{LimitFreeEnergies}}
In this section, we compute the limit when $\epsilon$ goes to $0$ of the free energies $\left(F^{(g)}_{\text{Top.Rec.}}(\epsilon)\right)_{g\geq 0}$ of the spectral curve:  
\bea \label{OldPara}x(z)&=&\frac{1}{2}\tan\left(\frac{\pi \epsilon}{2}\right)\left(z+\frac{1}{z}\right)\cr
y(z)&=&\frac{2}{\sin\left(\frac{\pi \epsilon}{2}\right)\left(1+\frac{1}{4}\tan^2\left(\frac{\pi \epsilon}{2}\right)\left(z+\frac{1}{z}\right)\right)\left(z-\frac{1}{z}\right)}\eea
Using the symplectic transformation:
\beq \label{ChangePara} \left(\td{x},\td{y}\right)=\left(\frac{x}{\tan \frac{\pi \epsilon}{2}},y\tan \frac{\pi \epsilon}{2}\right)\eeq
we obtain the equivalent spectral curve
\bea  \label{NewPara} \td{x}(z)&=&\frac{1}{2}\left(z+\frac{1}{z}\right)\cr
\td{y}(z)&=&\frac{2}{\cos\left(\frac{\pi \epsilon}{2}\right)\left(1+\frac{1}{4}\tan^2\left(\frac{\pi \epsilon}{2}\right)\left(z+\frac{1}{z}\right)\right)\left(z-\frac{1}{z}\right)}
\eea
In particular, since the symplectic transformation \eqref{ChangePara} is only a rescaling of $x$ and $y$ by a constant, the free energies computed from either \eqref{OldPara} or \eqref{NewPara} are the same for $g\geq 1$. The exceptional case $g=0$ provides the same result up to a trivial factor $\ln \tan \frac{\pi \epsilon}{2}$. Finally, the new parametrization \eqref{NewPara} admits a regular limit when $\epsilon\to 0$ given by:
\beq x_0(z)=\frac{1}{2}\left(z+\frac{1}{z}\right)\, \,,\,\, y_0(z)=\frac{2}{\left(z-\frac{1}{z}\right)}\eeq
which is exactly Legendre's spectral curve $y_0=\frac{1}{\sqrt{x_0^2-1}}$ whose free energies have recently been computed in \cite{IwakiAllCurves} (Cf. section $2.3.4$ and main theorem of \cite{IwakiAllCurves}). Thus, we get:
\bea \label{ResultsIwaki} F^{(0)}_{\text{Top.Rec.}}(\epsilon)+\ln \tan \frac{\pi\epsilon}{2}&\overset{\epsilon \to 0}{\to}&\ln 2\cr
F^{(1)}_{\text{Top.Rec.}}(\epsilon=0)&=&0\cr
F^{(g)}_{\text{Top.Rec.}}(\epsilon=0)&=&\frac{B_{2g}\left(4-2^{-(2g-2)}\right)}{2g(2g-2)}\,\, ,\,\, \forall\, g\geq 2\cr
&&\eea
where $\left(B_k\right)_{k\geq 0}$ are the Bernoulli numbers defined by:
\beq \frac{x}{e^x-1}=\sum_{k=0}^{\infty}B_k\frac{x^k}{k!}\eeq

\subsection{Convergence of  the asymptotic expansion $\underset{g=0}{\overset{\infty}{\sum}} F^{(g)}_{\text{Top.Rec.}}(\epsilon)n^{2-2g}$ when $\epsilon\to 0$\label{AsymptExpansionLimit}}
In the previous section, we have proved that:
\bea \label{ResultsIwaki22} F^{(0)}_{\text{Top.Rec.}}(\epsilon)+\ln \tan \frac{\pi\epsilon}{2}&\overset{\epsilon \to 0}{\to}&\ln 2\cr
F^{(1)}_{\text{Top.Rec.}}(\epsilon=0)&=&0\cr
F^{(g)}_{\text{Top.Rec.}}(\epsilon=0)&=&\frac{B_{2g}\left(4-2^{-(2g-2)}\right)}{2g(2g-2)}\,\, ,\,\, \forall\, g\geq 2\eea
However, the previous results are only term by term convergence results and may only be used to study formal series in $n^{-2g}$ but not directly in asymptotic expansions that we deal with in the present article. In order to prove that the limit remains valid in the sense of large $n$ asymptotic expansion as claimed in \eqref{ExactResults1}, we need to use results of \cite{BG2} rather than formal limit results found in \cite{EO}. Let us consider the integrals defined by (we remind that $a=\tan \frac{\pi\epsilon}{2}$):
\beq  S_{a}=\int_{[-1,1]^n} \Delta(u_1,\dots,u_n)^2e^{-n\underset{k=1}{\overset{n}{\sum}}\ln(1+a^2u_k^2)}du_1\dots du_n\eeq
which is trivially connected to $Z_n(a)$ by equation \eqref{CasRef}. Note in particular that the integration domain is $[-1,1]^n$ and thus is independent of $a$. On the contrary, the potential $V_a(u)=\ln (1+a^2u^2)$ depends continuously in $a$. In particular, $V_a$ converges uniformly on $[-1,1]$ to $V_0=0$ when $a$ goes to $0$. Moreover, for all $a\geq 0$, the potentials $V_a$ satisfy conditions of Proposition \ref{PropNew} and in particular the limiting eigenvalues density is always supported on a single interval. Therefore, we may apply results of \cite{BG2} stating that the large $n$ asymptotic expansion of $\ln S_a$ converges to the large $n$ asymptotic expansion of $\ln S_0$:
\beq \label{refeee}\sum_{g=0}^{\infty} \left(F^{(g)}_{\text{Top.Rec.}}(\epsilon)+\delta_{g,0}\ln \tan \frac{\pi\epsilon}{2}\right)n^{2-2g}\overset{\epsilon\to 0}{=}n^2\ln 2+ \sum_{g=2}^{\infty}\frac{B_{2g}\left(4-2^{-(2g-2)}\right)}{2g(2g-2)}n^{2-2g}\eeq
 Eventually, the connection between $Z_n(a)$ to $S_a$ given by \eqref{CasRef} allows to extend the result to the large $n$ asymptotic expansions of $\ln Z_n(a)$ as claimed in \eqref{ExactResults1}.  

\begin{remark}The case $a=0$ only appears special at first sight because the definition of $Z_n(a)$ (See \eqref{RealInt2}) is taken as integrals over $[-a,a]$ for which the value $a=0$ looks ill-defined. But after a trivial transformation to integrals over $[-1,1]$ given by \eqref{CasRef} (i.e. transforming $Z_n(a)$ to $S_a$), it becomes obvious that $a=0$ is no different than any $a>0$ regarding the application of results of \cite{BG2}. Therefore, the asymptotic expansion of $\ln S_a$ converges to the asymptotic expansion of $\ln S_0$ when $a$ tends to $0$ thus giving \eqref{refeee} and finally \eqref{ExactResults1} using \eqref{CasRef}.
\end{remark}

\section{The Eynard-Orantin topological recursion\label{AppendixTopRec}}

In this section we briefly review the formalism of the topological recursion as presented in \cite{EO}. More general versions of the topological recursion can be found in the literature but we restrict ourselves to the original simpler version of \cite{EO} that is sufficient for the purposes of this article. Let us start by the definition of a spectral curve:

\begin{definition}[Spectral curve, branchpoints, normalized bi-differential] A spectral curve is the data of two meromorphic functions $(x(z),y(z))$ on a Riemann surface $\Sigma$ of genus $\genus$. This is equivalent to the data of a polynomial $P$ such that $P(x,y)=0$ and therefore to an algebraic equation between $x$ and $y$. When the genus $\genus$ of $\Sigma$ is strictly positive, we complete the data of the spectral curve with the choice of a basis of homology cycles $\left(\mathcal{A}_j,\mathcal{B}_j\right)_{1\leq j\leq \genus}$ such that:
\beaa \forall \,1\leq j,k\leq \genus\,&:&\, \mathcal{A}_j\cap\mathcal{A}_k=\mathcal{B}_j\cap\mathcal{B}_k=\emptyset\cr
 \forall \,1\leq j,k\leq \genus\,&:&\, \mathcal{A}_j\cap\mathcal{B}_k=\delta_{j,k}
\eeaa
Then it follows from standard results of algebraic geometry that there exists a unique symmetric bi-differential $B(z_1,z_2)$ (sometimes called ``Bergman kernel'') such that:
\begin{itemize}\item $B$ is holomorphic on $\Sigma\times\Sigma$ except at coinciding points where it behaves like:
\beqq B(z_1,z_2)=\frac{dz_1dz_2}{(z_1-z_2)^2} +\text{regular}_{z_1\to z_2}\eeqq
\item $B$ is normalized on the basis of cycles $\left(\mathcal{A}_j,\mathcal{B}_j\right)_{1\leq j\leq \genus}$ in the following way:
\beqq \oint_{\mathcal{A}_j}B(z_1,z_2)=0 \text{ for all } 1\leq j\leq \genus\eeqq
\end{itemize} 
The branchpoints $(a_j)_{1\leq j\leq R}$ (with $R\geq 1$) of the spectral curve are the points where $dx$ vanishes. The spectral curve is said ``regular'' if the branchpoints are simple zeros of $dx$. When the spectral curve is regular, we can define locally around each branchpoint a holomorphic involution $z\mapsto \bar{z}$ such that $x(z)=x(\bar{z})$. 
\end{definition} 

In this paper we will only restrict ourselves to the case of regular spectral curves since the situation is much more complicated when the curve is not regular. We remark that when the spectral curve is regular and of genus $0$, then there exists a global parametrization $(x(z),y(z))$ with $z\in\mathbb{C}\cup\{\infty\}$ of the spectral curve. Moreover, the holomorphic involution $z\mapsto \bar{z}$ is defined globally on the spectral curve and the normalized bi-differential $B$ is explicitly given by $B(z_1,z_2)=\frac{dz_1dz_2}{(z_1-z_2)^2}$. We now have all the ingredients to define the correlators and free energies associated to a spectral curve.  

\begin{definition}[Definition 4.2 of \cite{EO}] For $g \geq 0$ and $n \geq 1$, the Eynard-Orantin differentials (known also as ``correlation functions'' or ``correlators'') $\omega^{(g)}_{n}(z_{1},\dots,z_{n})$ of type $(g,n)$ associated to the spectral curve $(x(z),y(z))$ are defined by the following recursive relations:
\bea \label{eqtoprec}
\omega^{(0)}_{1}(z_{1}) & = & (y(z_{1})-y(\bar{z}_1))dx(z_{1}) \\
\omega^{(0)}_{2}(z_{1},z_{2}) & = & B(z_1,z_2), \\ 
\omega^{(g)}_{n+1}(z_{0},z_{1},\dots,z_{n}) & = & \underset{i=1}{\overset{R}{\sum}} \underset{z\to a_i}{\Res} \,K(z_{0},z) \Big[ \omega^{(g-1)}_{n+1}(z,\bar{z},z_{1},\dots,z_{n}) \\
& & + \sum'_{\substack{g_{1} + g_{2} = g \cr 
I \sqcup J = \{1,\dots,n\} }} \omega^{(g_{1})}_{1+|I|}(z,z_{I})\omega^{(g_{2})}_{1+|J|}(\bar{z},z_{J}) \Big].  \nonumber
\eea
Here 
\beq
K(z_{0},z) = \frac{1}{2}\frac{\int^{\bar{z}}_{z} \omega^{(0)}_{2}(\cdot, z_{0})}{(y(z)-y(\bar{z}))dx(z)}
\eeq
is called the recursion kernel, and the $\,'$ in the last line of \eqref{eqtoprec} means that the cases $(g_{1}, I) = (0, \emptyset)$ and $(g_{2}, J) = (0, \emptyset)$ must be excluded from the sum. 
\end{definition}

The Eynard-Orantin differentials $\omega^{(g)}_{n}$'s are meromorphic multi-differentials on $\Sigma^n$ and are known to be holomorphic except at the branchpoints if $(g,n) \ne (0,1), (0,2)$. In \cite{EO}, the authors also introduced free energies (also called ``symplectic invariants'') $\left(F^{(g)}\right)_{g\geq 0}$ defined by: 

\begin{definition}[Definition 4.3 of \cite{EO}] The $g^{\text{th}}$ symplectic invariant $F^{(g)}$ associated to the spectral curve $(x(z),y(z))$ is defined by: 
\beqq
F^{(g)} = \frac{1}{2-2g}\underset{i=1}{\overset{R}{\sum}} \Res_{z \to a_i} \Phi(z) \omega^{(g)}_{1}(z) \quad \text{for $g \geq 2$}
\eeqq
where 
\beqq
\Phi(z) = \int^{z}_{z_{o}} y(\tilde{z})dx(\tilde{z}) \quad\text{($z_{o}$ is a generic point)}.
\eeqq
$F^{(0)}$ and $F^{(1)}$ are defined with specific formulas that can be found in \cite{EO}.
\end{definition}

Note that this definition extends to the case $n=0$ (with the identification $\omega_0^{(g)}=F^{(g)}$) the following property:
\beqq  \omega_n^{(g)}(z_1,\dots,z_n) = \frac{1}{2-2g-n}\,\underset{i=1}{\overset{R}{\sum}} \Res_{z \to a_i} \Phi(z) \omega^{(g)}_{n+1}(z,z_1,\dots,z_n) \quad \text{for $g \geq 0$ and $n\geq 0$}\eeqq
Note also that the Eynard-Orantin differentials or the symplectic invariants do not depend on the choice of parametrization $(x(z),y(z))$. Eventually as suggested by their name, the symplectic invariants $\left(F^{(g)}\right)_{g\geq 0}$ are invariant under transformations of the spectral curve $(x,y)\to (\td{x},\td{y})$ such that $dx\wedge dy=d\td{x}\wedge d\td{y}$, i.e. transformations that preserve the symplectic form $dx\wedge dy$. Note that the only non-trivial case is the exchange $(x,y)\rightarrow (y,-x)$ for which the invariance of the symplectic invariants is highly non-trivial and controversial. In this article we only used trivial symplectic transformations. We remind that this symplectic invariance property does not hold in general for the Eynard-Orantin differentials $\omega_n^{(g)}$ but only for the free energies.

\end{appendix}

\end{document}